\documentclass[a4paper,11pt]{article}

\usepackage{amsmath}
\usepackage{amsfonts}
\usepackage{graphicx}
\usepackage{bm}

\usepackage{enumerate}
\usepackage{color}
\usepackage{float}
\usepackage{multirow}

\usepackage[top=2.54cm,bottom=2.54cm, left=2.54cm, right=2.54cm]{geometry}
\usepackage{amssymb}
\usepackage[sort,comma]{natbib}
\usepackage[colorlinks=true, allcolors=blue]{hyperref}
\usepackage{rotating}
\usepackage{etoolbox}
\usepackage{tikz}
\usepackage{rotating}
\usepackage{changepage}

\newrobustcmd*{\mysquare}[1]{\tikz{\filldraw[draw=#1,fill=#1] (0,0)
rectangle (0.2cm,0.2cm);}}

\newrobustcmd*{\mycircle}[1]{\tikz{\filldraw[draw=#1,fill=#1] (0,0) circle [radius=0.1cm];}}

\newrobustcmd*{\mytriangle}[1]{\tikz{\filldraw[draw=#1,fill=#1] (0,0) --
(0.2cm,0) -- (0.1cm,0.2cm);}}

\title{Flexible non-parametric regression models for compositional response data with zeros}
\author{Michail Tsagris, Abdulaziz Alenazi and Connie Stewart\\
\\
Department of Economics, University of Crete, Greece  \\
\href{mailto:mtsagris@uoc.gr}{mtsagris@uoc.gr} \\
Department of Mathematics, College of Science, Northern Border University,  Saudi Arabia \\ \href{mailto:a.alenazi@nbu.edu.sa}{a.alenazi@nbu.edu.sa} \\
Department of Mathematics and Statistics, University of New Brunswick, Saint John, Canada \\ \href{mailto:Connie.Stewart@unb.ca}{Connie.Stewart@unb.ca} \\
}

\begin{document}

\maketitle

\begin{center}
\textbf{Abstract}
\end{center}
Compositional data arise in many real-life applications and versatile methods for properly analyzing this type of data in the regression context are needed. When parametric assumptions do not hold or are difficult to verify, non-parametric regression models can provide a convenient alternative method for prediction. To this end, we consider an extension to the classical $k$--$NN$ regression, termed $\alpha$--$k$--$NN$ regression, that yields a highly flexible non-parametric regression model for compositional data through the use of the $\alpha$-transformation. Unlike many of the recommended regression models for compositional data, zeros values (which commonly occur in practice) are not problematic and they can be incorporated into the proposed models without modification. Extensive simulation studies and real-life data analyses highlight the advantage of using these non-parametric regressions for complex relationships between the compositional response data and Euclidean predictor variables. Both suggest that $\alpha$--$k$--$NN$ regression can lead to more accurate predictions compared to current regression models which assume a, sometimes restrictive, parametric relationship with the predictor variables. In addition, the $\alpha$--$k$--$NN$ regression, in contrast to current regression techniques, enjoys a high computational efficiency rendering it highly attractive for use with large scale, massive, or big data. 

\section{Introduction}
Non-negative multivariate vectors with variables (typically called components) conveying only relative information are referred to as compositional data. When the vectors are normalized to sum to 1, their sample space is the standard simplex given below
\begin{eqnarray} \label{simplex}
\mathbb{S}^{D-1}=\left\lbrace(u_1,...,u_D)^\top \bigg\vert u_i \geq 0,\sum_{i=1}^Du_i=1\right\rbrace, 
\end{eqnarray}
where $D$ denotes the number of components. 

Examples of compositional data may be found in many different fields of study and the extensive scientific literature that has been published on the proper analysis of this type of data is indicative of its prevalence in real-life applications\footnote{For a substantial number of specific examples of applications involving compositional data see \citep{tsagris2020}.}. It is perhaps not surprising, given the widespread occurrence of this type of data, that many compositional data analysis applications involve covariates. In sedimentology, for example, samples were collected from an Arctic lake and the change in their chemical composition across different water depths was of interest \citep{compositions2018}. This data set is analyzed in Section \ref{data} using our proposed methodology along with several other data sets. These include compositional glacial data, household consumption expenditures data, data on the concentration of chemical elements in samples of soil, data on morphometric measurements of fish, as well as electoral, pollution and power data, all of which are associated with some covariates. In addition to these examples, the literature cites numerous other applications of compositional regression analysis. In economics, \cite{morais2018} linked market shares to some independent variables, while in political sciences the percentage of votes of each candidate were linked to some relevant predictor variables \citep{katz1999, tsagris2018b}. Finally, in the field of bioinformatics compositional data techniques have been used for analysing microbiome data \citep{xia2013, chen2016, shi2016}.     

The need for valid regression models for compositional data in practice has led to several developments in this area, many of which have been proposed in recent years. The first regression model for compositional response data was developed by \cite{ait2003}, commonly referred to as Aitchison's model, and was based on the additive log-ratio transformation defined in Section \ref{trans}. Dirichlet regression was applied to compositional data in \cite{gueorguieva2008, hijazi2009, melo2009}. The additive log-ratio transformation was again used by \cite{tolosana2009} while \cite{egozcue2012} extended Aitchison's regression model by using an isometric log-ratio transformation but, instead of employing the usual Helmert sub-matrix, \cite{egozcue2012} chose a different orthogonal matrix that is compositional data dependent. 

A drawback of the aforementioned regression models is their inability to handle zero values directly and, consequently, a few models have recently been proposed to address the zero problem. In particular, \cite{scealy2011} transformed the compositional data onto the unit hyper-sphere and introduced the Kent regression which treats zero values naturally. Spatial compositional data with zeros were modelled in \cite{leininger2013} from the Bayesian stance. Alternative regression models in the field of econometrics and applicable when zero values are present are discussed in \cite{murteira2016}. In \cite{tsagris2015a}, a regression model that minimizes the Jensen-Shannon divergence was proposed while in \cite{tsagris2015b}, $\alpha$-regression (a generalization of Aitchison's log-ratio regression) was introduced, and both of these approaches are compatible with zeros. An extension to Dirichlet regression allowing for zeros was developed by \cite{tsagris2018b} and referred to as zero adjusted Dirichlet regression.

The contribution of this paper is an extension of these classical non-parametric approaches for application to compositional data through the utilization of the $\alpha$--transformation. Specifically, the proposed $\alpha$--$k$--$NN$ regression for compositional data link the predictor variables in a non-parametric, non-linear fashion, thus allowing for more flexibility. The models have the potential to provide a better fit to the data compared to conventional models and yield improved predictions when the relationships between the compositional and the non-compositional variables are complex. Furthermore, in contrast to other non-parametric regressions such as projection pursuit, applicable to log-ratio transformed compositional data (see \cite{friedman1981}), the two proposed methods allow for zero values in the data. Finally, a significant advantage of $\alpha$--$k$--$NN$ regression in particular is its high computational efficiency compared to all current regression techniques, even when the sample sizes number hundreds of thousands or even millions of observations. Functions to carry out both $\alpha$--$k$--$NN$ regression are provided in the \textit{R} package \textit{Compositional}.

The paper is structured as follows: Section \ref{trans} describes relevant transformations and regression models for compositional data, while in Section \ref{aknnreg}, the proposed $\alpha$--$k$--$NN$ regression model is introduced. We examine the advantages and limitations of these models through simulation studies, implemented in Section \ref{simulations}, and real-life data sets are presented in Section \ref{data}. Finally, concluding remarks are provided in Section \ref{conclusion}.

\section{Compositional data analysis: transformations and regression models}
\label{trans}
In this section some preliminary definitions and methods in compositional data analysis relevant to the work in this paper are introduced. Specifically, two commonly used log-ratio transformations, as well as a more general $\alpha$--transformation are defined, followed by some existing regression models for compositional regression.

\subsection{Transformations}
\subsubsection{Additive log-ratio transformation}
\cite{ait1982} suggested applying the additive log-ratio (alr) transformation to compositional data prior to using standard multivariate data analysis techniques. Let ${\bf u}=\left(u_1,\ldots, u_D\right)^\top \in \mathbb{S}^{D-1}$, the alr transformation is given by 
\begin{eqnarray} \label{alr}
\mathbf{v} =\left\{v_i \right \}_{i=1,\ldots,D-1} = \left \{ \log \frac{u_i}{u_1} \right \}_{i=2,\ldots,D},
\end{eqnarray}
where ${\bf v}=\left(v_1,\ldots,v_{D-1}\right) \in \mathbb{R}^{D-1}$. Note that the common divisor, $u_1$, need not be the first component and was simply chosen for convenience. The inverse of Equation (\ref{alr}) is given by
\begin{eqnarray} \label{alrinv}
\mathbf{u} =\left\{u_i \right \}_{i=1,\ldots,D} = 
\left \lbrace
\begin{array}{cc}
\frac{1}{1 + \sum_{j=2}^De^{{v}_j}},  & i = 1 \\
\frac{e^{{v}_{i-1}}}{1 + \sum_{j=2}^De^{{v}_j}},  & i = 2, \ldots, D 
\end{array}
\right \rbrace.
\end{eqnarray}

\subsubsection{Isometric log-ratio transformation}
An alternative transformation proposed by \cite{ait1983} is the centred log-ratio (clr) transformation defined as 
\begin{eqnarray} \label{clr}
\mathbf{y} = \left \{y_i \right \}_{i=1,\ldots,D} = \left \{ \log{\frac{u_i}{g\left({\bf u}\right)}} \right \}_{i=1,\ldots,D,}
\end{eqnarray}
where $g\left({\bf u}\right)=\prod_{j=1}^Du_j^{1/D}$ is the geometric mean. The inverse of Equation (\ref{clr}) is given by 
\begin{eqnarray} \label{clrinv} 
\mathbf{u} =\left\{u_i \right \}_{i=1,\ldots,D} = \left\lbrace\frac{e^{y_i}}{\sum_{j=1}^De^{y_j}} \right\rbrace_{i=1, \ldots, D} = \mathcal{C}\left\lbrace e^{{\bf y}} \right\rbrace,
\end{eqnarray}
where $\mathcal{C}\left\lbrace . \right\rbrace$ denotes the closure operation, or normalization to the unity sum.

The clr transformation in Equation (\ref{clr}) was proposed in the context of principal component analysis and its drawback is that $\sum_{i=1}^D y_i=0$, so essentially the problem of the unity sum constraint is replaced by the problem of the zero sum constraint. In order to address this issue, \cite{ilr2003} proposed multiplying Equation (\ref{clr}) by the $(D-1) \times D$ Helmert sub-matrix $\bf H$ \citep{helm1965,dryden1998,le1999}, an orthogonal matrix with the first row omitted, which results in what is called the isometric log-ratio (ilr) transformation
\begin{eqnarray} \label{ilr}
\mathbf{z}_{0} = {\bf H}\mathbf{y},
\end{eqnarray}
where $\mathbf{z}_{0} = \left(z_{01},\ldots, z_{0D-1}\right)^\top \in \mathbb{R}^{D-1}$. Note that any orthogonal matrix which preserves distances would also be appropriate in place of $\bf H$ \citep{tsagris2011}. The inverse of Equation (\ref{ilr}) is 
\begin{eqnarray} \label{ilrinv}
\mathbf{u}_0 =\mathcal{C}\left\lbrace e^{{\bf H}^\top{\bf z}_0} \right\rbrace.
\end{eqnarray}

\subsubsection{$\alpha$--transformation}
The main disadvantage of the above transformations is that they do not allow zero values in any of the components, unless a zero value imputation technique (see \cite{martin2003}, for example) is first applied. This strategy, however, can produce regression models with predictive performance worse than regression models that handle zeros naturally \citep{tsagris2015a}. When zeros occur in the data, the power transformation introduced by \cite{ait2003}, and subsequently modified by \cite{tsagris2011}, may be used. Specifically, \cite{ait2003} defined the power transformation as
\begin{eqnarray} \label{alpha}
{\bf w}_{\alpha}=\{w_i\}_{\alpha, i=1, \ldots , D}=\mathcal{C}\left\lbrace u_i^{\alpha}\right\rbrace
\end{eqnarray}
and \cite{tsagris2011} defined the $\alpha$-transformation, based on Equation (\ref{alpha}), as
\begin{eqnarray} \label{isoalpha}
{\bf z}_{\alpha}=\frac{1}{\alpha}{\bf H}\left(D{\bf w}_{\alpha}-{\bf 1}_D\right), 
\end{eqnarray} 
where ${\bf H}$ is the Helmert sub-matrix and ${\bf 1}_D$ is the $D$-dimensional vector of 1s. While the power transformed vector ${\bf w}_{\alpha}$ in Equation (\ref{alpha}) remains in the simplex $\mathbb{S}^{D-1}$, ${\bf z}_{\alpha}$ in Equation (\ref{isoalpha}) is mapped onto a subset of $\mathbb{R}^{D-1}$. Furthermore, as $\alpha \rightarrow 0$, Equation (\ref{isoalpha}) converges to the ilr transformation in Equation (\ref{ilr}) \citep{tsagris2016}. For convenience purposes, $\alpha$ is generally taken to be between $-1$ and $1$, but when zeros occur in the data, $\alpha$ must be restricted to be strictly positive. The inverse of ${\bf z}_\alpha$ is
\begin{eqnarray}  \label{alphaInverse}
{\bf u}_{\alpha}  = \mathcal{C}\left\lbrace \left( \alpha {\bf H}^\top{\bf z}_{\alpha}+ {\bf 1}_D \right)^{1/\alpha} \right\rbrace.
\end{eqnarray}

\cite{tsagris2011} argued that while the $\alpha$-transformation did not satisfy some of the properties that \cite{ait2003} deemed important, this was not a downside of this transformation as those properties were suggested mainly to fortify the concept of log-ratio methods. \cite{scealy2014} also questioned the importance of these properties and, in fact, showed that some of them are not actually satisfied by the log-ratio methods that they were intended to justify.

\subsection{Regression models for compositional data} 

\subsubsection{Additive and isometric log-ratio regression models}
Let $\bf V$ denote the response matrix with $n$ rows containing alr transformed compositions. $\bf V$ can then be linked to some predictor variables $\bf X$ via
\begin{eqnarray} \label{aitreg}
{\bf V} = {\bf XB} + {\bf E},
\end{eqnarray}
where ${\bf B}=\left(\pmb{\beta}_2, \ldots, \pmb{\beta}_{D}\right)$ is the matrix of coefficients, $\bf X$ is the design matrix containing the $p$ predictor variables (and the column of ones) and $\bf E$ is the residual matrix. Referring to Equation (\ref{alr}), Equation (\ref{aitreg}) can be re-written as  
\begin{eqnarray} \label{regalr}
\log\left(\frac{u_{ij}}{u_{1j}}\right)={\bf X}^\top_j\pmb{\beta}_i \Leftrightarrow
\log{u_{ij}}=\log{u_{1j}}+{\bf X}^\top_j\pmb{\beta}_i, \ \ i=2,\ldots, D, \  j=1,\ldots, n,
\end{eqnarray}
where ${\bf X}^\top_j$ denotes the $j$th row of $\bf{X}$.  Equation (\ref{regalr}) can be found in \cite{tsagris2015b} where it is shown that the alr regression (\ref{aitreg}) is in fact a multivariate linear regression in the logarithm of the compositional data with the first component (or any other component) playing the role of an offset variable; an independent variable with coefficient equal to $1$. 

Regression based on the ilr tranformation (ilr regression) is similar to alr regression and is carried out by substituting $\bf V$ in Equation (\ref{aitreg}) by ${\bf Z}_0$ in Equation (\ref{ilr}). The fitted values for both the alr and ilr transformations are the same and are therefore generally back transformed onto the simplex using the inverse of the alr transformation in Equation (\ref{alrinv}) for ease of interpretation.

\subsubsection{Kullback-Leibler divergence based regression}
\cite{murteira2016} estimated the $\pmb{\beta}$ coefficients via minimization of the Kullback-Leibler divergence 
\begin{eqnarray} \label{klreg}
\min_{\pmb{\beta}} \sum_{j=1}^n{\bf u}_j^\top\log{\frac{{\bf y}_j}{\hat{\pmb{\mu}}_j}}=
\max_{\pmb{\beta}} \sum_{j=1}^n{\bf u}_j^\top\log{\hat{\pmb{\mu}}_j}, 
\end{eqnarray}
where $\hat{\pmb{\mu}}_j=\left(\hat{\mu}_{j1}, \ldots, \hat{\mu}_{jD}\right)^T$ is 
\begin{eqnarray}  \label{regalpha}
\hat{\mu}_{ij}=
\left\lbrace
\begin{array}{cc} 
\frac{1}{1+\sum_{k=2}^De^{{\bf x}_j^T\pmb{\beta}_k}} & \text{if} \ \ i=1 \\
\frac{e^{{\bf x}_j^T\pmb{\beta}_i}}{1+\sum_{k=2}^De^{{\bf x}^T\pmb{\beta}_k}} & \text{for} \ \ i=2,...,D,
\end{array} 
\right\rbrace
\end{eqnarray}
with  
$\pmb{\beta}_i=\left(\beta_{0i},\beta_{1i},...,\beta_{pi} \right)^T$,  $i=1,\ldots,d$. \citep{tsagris2015a,tsagris2015b} 

The regression model in Equation (\ref{klreg}), also referred to as Multinomial logit regression, will be denoted by Kullback-Leibler Divergence (KLD) regression throughout the rest of the paper. KLD regression is a semi-parametric regression technique, and, unlike alr and ilr regression, can handle zeros naturally.

\section{The $\alpha$--$k$--$NN$ Regression Model} \label{aknnreg}
Our proposed $\alpha$--$k$--$NN$ regression model extend the well-known $k$--$NN$ regression to the compositional data setting. Unlike previous approaches, the models are more flexible (due to the $\alpha$ parameter), entirely non-parametric and allow for zeros, thus filling an important gap in the compositional data analysis literature

In general terms, to predict the response value corresponding to a new vector of predictor values (${\bf x}_{new}$), the $k$--$NN$ algorithm first computes the Euclidean distances from ${\bf x}_{new}$ to the observed predictor values ${\bf x}$. $k$--$NN$ regression works by selecting the response values coinciding with the observations with the $k$--smallest distances between ${\bf x}_{new}$ and the observed predictor values, and then averaging those response values, using the sample mean or median, for example. The proposed $\alpha$--$k$--$NN$ regression algorithm relies on an extension of the sample mean to the sample Fr{\'e}chet mean, described below.

\subsection{The Fr{\'e}chet mean}
The Fr{\'e}chet mean \citep{pennec1999} on a metric space $\left(\mathbb{M}, \text{dist} \right)$, with the distance between $p,q\in \mathbb{M}$ given by $\text{dist}\left(p,q\right)$, is defined by   
$\text{argmin}_{h \in \mathbb{M}}E_U\left[\text{dist}\left(U, h\right)^2 \right]$ and $\text{argmin}_{h \in \mathbb{M}}\sum_{j=1}^n\text{dist}\left(u_j, h\right)^2$ in the population and finite sample cases, respectively. The Fr{\'e}chet mean in the compositional data setting for a sample size $n$ is the argument $\pmb{\gamma}=\left(\gamma_1,\ldots,\gamma_D\right)^T$ which minimizes the following expression in \cite{tsagris2011}
\begin{eqnarray*}
\pmb{\mu}_{\alpha}=\sum_{j=1}^n\left[\sum_{i=1}^D\left( \frac{u_{ij}^{\alpha}}{\sum_{k=1}^Du_{kj}^{\alpha}}-\gamma_i \right)^2\right],
\end{eqnarray*}
where $\gamma_i=\frac{\mu_i^{\alpha}}{\sum_{k=1}^D\mu_k^{\alpha}}$.
Setting $\partial\pmb{\mu}_{\alpha}/\partial \gamma_i=0$, the minimization occurs at
\begin{eqnarray*}
\hat{\gamma_i}=\frac{1}{n}\sum_{j=1}^n\left(\frac{u_{ij}^{\alpha}}{\sum_{k=1}^Du_{kj}^{\alpha}} \right),
\end{eqnarray*}
and thus 
\begin{eqnarray} \label{frechet}
\hat{\bm{\mu}}_{\alpha}\left({\bf u}\right)=\mathcal{C}\left\lbrace\left\lbrace\left[\frac{1}{n}\sum_{j=1}^n\left(\frac{u_{ij}^\alpha}{\sum_{k=1}^Du_{kj}^\alpha}\right)\right]^{1/\alpha}\right\rbrace_{i=1,\ldots,D}\right\rbrace = \mathcal{C}\left\lbrace \left\lbrace \left(\sum_{j=1}^n u^{\alpha}_ {ij}\right)^{1/\alpha}\right \rbrace_{i=1,\ldots,D} \right \rbrace.
\end{eqnarray}

\cite{kendall2011} showed that the central limit theorem applies to Fr{\'e}chet means defined on manifold valued data and the simplex space is an example of a manifold \citep{pantazis2019}. Another nice property of the Fr{\'e}chet mean is that it is absolutely continuous as $\alpha$ tends to zero, in the limiting case, Equation (\ref{frechet}) converges to the closed geometric mean, $\hat{\bm{\mu}}_0$ defined below and in \cite{ait1989}. That is, 
\begin{eqnarray*}
\lim_{\alpha \rightarrow 0}{\hat{\bm{\mu}}_{\alpha}\left({\bf u}\right)} \rightarrow \hat{\bm{\mu}}_0\left({\bf u}\right) = \mathcal{C}\left\lbrace  \left\lbrace \left(\prod_{j=1}^n u_{ij}\right)^{1/n}  \right \rbrace_{i=1,\ldots,D} \right \rbrace.
\end{eqnarray*}

\subsection{The $\alpha$--$k$--$NN$ regression}

When the response variables, ${\bf u}_1,{\bf u}_2,\ldots,{\bf u}_n$, represent compositional data, the $k$--$NN$ algorithm can be applied to the transformed data in a straightforward manner, for a specified transformation appropriate for compositional data. 
For added flexibility and to be able to handle compositional data with zeros directly, we propose extending $k$--$NN$ regression using the power transformation in Equation (\ref{alpha}) combined with the Fr{\'e}chet mean in Equation (\ref{frechet}). Specifically, in $\alpha$--$k$--NN regression, the predicted response value corresponding to ${\bf x}_{new}$ is then
\begin{eqnarray} \label{aknnfit}
\hat{\bm{\mu}}_{\alpha}\left({\bf u}_j|j\in \mathcal{A}\right)=\hat{\bm{\mu}}_{\alpha, new} = \mathcal{C}\left\lbrace\left\lbrace\left(\sum_{j \in \mathcal{A}}u_{ij}^\alpha\right)^{1/\alpha}\right\rbrace_{i=1,\ldots , D} \right \rbrace,
\end{eqnarray}
where $\mathcal{A}$ denotes the set of $k$ observations, the $k$ nearest neighbours. In the limiting case of $\alpha=0$, the predicted response value is then
\[
\hat{\bm{\mu}}_0\left({\bf u}_j|j\in \mathcal{A}\right)=
\hat{\bm{\mu}}_{new} = \mathcal{C}\left\lbrace \left\lbrace \left(\prod_{j \in \mathcal{A}}u_{ij}\right)^{1/k} \right\rbrace_{i=1,\ldots, D}\right\rbrace.
\]
It is interesting to note that the limiting case also results from applying the clr transformation to the response data, taking the mean of the relevant $k$ transformed observations and then back transforming the mean using Equation (\ref{clrinv}). To see this, let  
\begin{eqnarray*}
\hat{\bf y}_{new} &=& \left\lbrace \frac{1}{k}\sum_{j \in \mathcal{A}}{ y}_{ij} \right\rbrace_{i=1,\ldots,D} = \left\lbrace\frac{1}{k}\sum_{j \in \mathcal{A}}\log{\frac{u_{ij}}{g\left({\bf u}_j\right)}} \right\rbrace_{i=1,\ldots,D} \\
\text{and hence} \\
\hat{\bf u}_{new} &=& 
\mathcal{C}\left\lbrace\left\lbrace \frac{e^{\hat{y}_{new, i}}}{\sum_{l=1}^De^{\hat{ y}_{new, i}}} \right\rbrace_{i=1,\ldots,D} \right\rbrace = 
\mathcal{C}\left\lbrace\left\lbrace \frac{e^{\frac{1}{k}\sum_{j \in \mathcal{A}}\log{\frac{{ u}_{ij}}{g\left({\bf u}_j\right)}}}}{\sum_{l=1}^De^{\frac{1}{k}\sum_{j \in \mathcal{A}}\log{\frac{{u}_{jj}}{g\left({\bf u}_j\right)}}}} \right\rbrace_{i=1,\ldots,D} \right\rbrace \\
&=& \mathcal{C}\left\lbrace\left\lbrace \frac{\prod_{j \in \mathcal{A}}{u}_{ij}^{1/k}}{\sum_{l=1}^D\prod_{j \in \mathcal{A}}{u}_{lj}^{1/k}} \right\rbrace_{i=1,\ldots,D} \right\rbrace  = \mathcal{C}\left\lbrace \left\lbrace \prod_{j \in \mathcal{A}} u_{ij}^{1/k} \right\rbrace_{i=1,\ldots, D}\right\rbrace. 
\end{eqnarray*}



\begin{figure}[!ht]
\centering
\begin{tabular}{cccc}
\includegraphics[scale = 0.43, trim = 30 40 10 10]{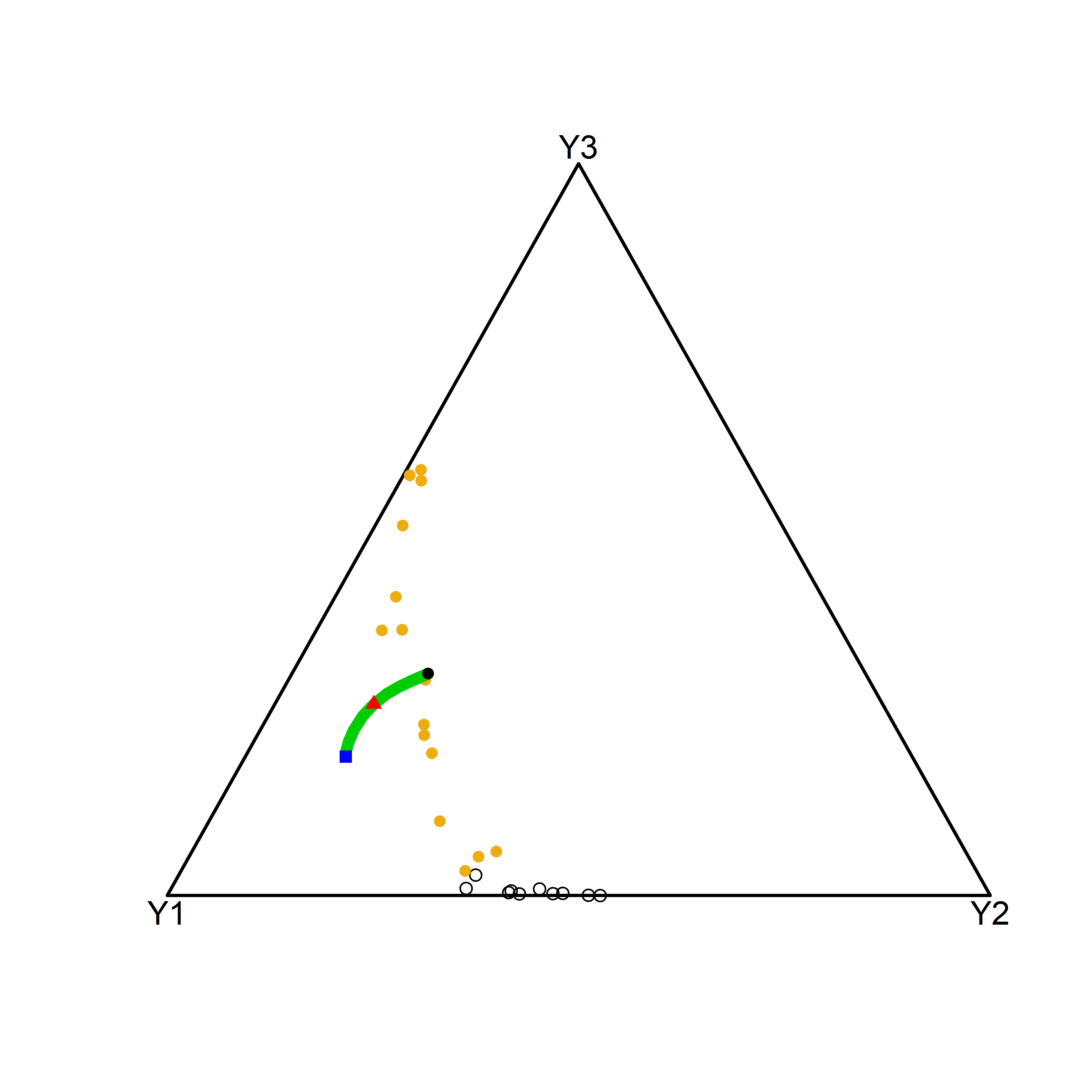}  &
\includegraphics[scale = 0.43, trim = 30 40 30 10]{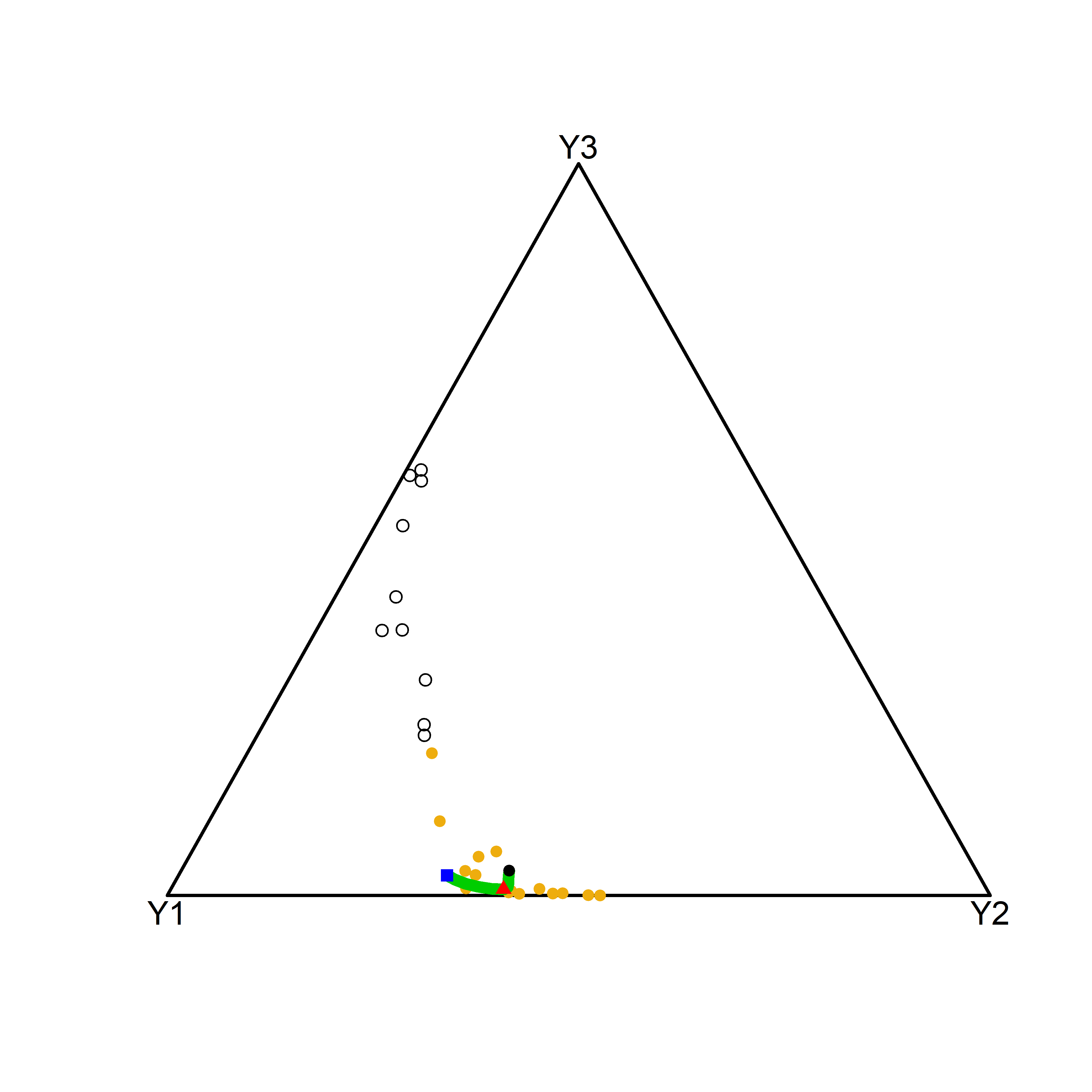}  \\
\textbf{(a)} Path of $\hat{\mu}_{\alpha}$ with 15 neighbours.  &  \textbf{(b)} Path of $\hat{\mu}_{\alpha}$ with 15 neighbours. \\
\multicolumn{2}{c}{\includegraphics[scale = 0.43, trim = 30 40 10 0]{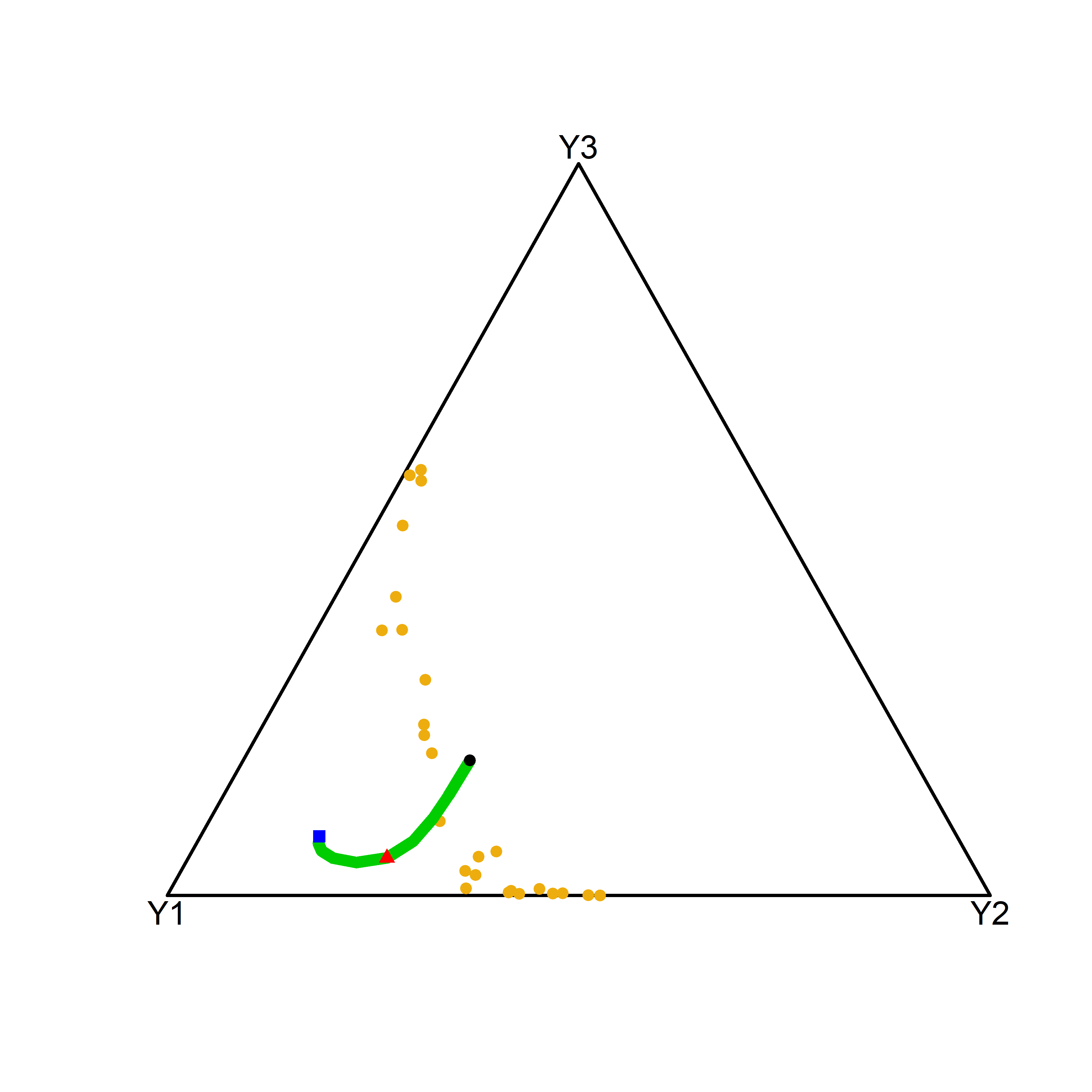} } \\ \multicolumn{2}{c}{ \textbf{(c)} Path of $\hat{\mu}_{\alpha}$ with 25 neighbours.}
\end{tabular}
\caption{\textbf{Effect of $\pmb{\alpha}$ and $\bf k$}. The symbols are as follows: \mysquare{blue} = Fr{\'e}chet mean with $\alpha=-1$, \mytriangle{red} = Fr{\'e}chet mean with $\alpha=0$ and \mycircle{black} = Fr{\'e}chet mean with $\alpha=1$. The dashed green curve \textcolor{green}{\textbf{-}} shows the path of all Fr{\'e}chet means starting with $\alpha=-1$ up to $\alpha=1$. The golden circles indicate the set of observations used to compute the Fr{\'e}chet mean. \label{effect}}
\end{figure}

A visual representation of the added flexibility provided by $\alpha$ on the Fr{\'e}chet mean, in addition to the effect of $k$, is given in Figure \ref{effect}.  Figure \ref{effect} shows how the Fr{\'e}chet mean varies when $\alpha$ moves from $-1$ to $1$ and different nearest neighbours are used. In Figures \ref{effect}(a) and \ref{effect}(b) two different sets of 15 neighbours are used, whereas in Figure \ref{effect}(c) all 25 nearest neighbours are used. The closed geometric mean (i.e. Fr{\'e}chet mean with $\alpha=0$) may lay outside the body of those observations. The Fr{\'e}chet mean on the contrary offers a higher flexibility, which in conjunction with the number of nearest neighbour $k$ yields estimates that can lie within the region of the selected set of compositional observations as visualised in Figure \ref{effect}. 

\subsection{Theoretical remarks}
The general family of nearest regression estimators, under weak regularity conditions, were shown to be uniformly consistent with probability one and the corresponding rate of convergence is near-optimal \citep{cheng1984}. More recently, \cite{jiang2019} proved the non-asymptotic uniform rates of consistency for the $k$--$NN$ regression\footnote{For more asymptotic results, see the references cited within  \cite{jiang2019}.}. The proof of consistency of the $\alpha$--$k$--$NN$ regression estimator falls within the work of \cite{lian2011} who dealt with the case of the response variable belonging in a separable Hilbert space. \cite{lian2011} investigated the rates of strong (almost sure) convergence of the $k$--$NN$ estimate under finite moment conditions and exponential tail condition on the noises. Recall that the simplex in Equation (\ref{simplex}) is a $D-1$ Hilbert space \citep{pawlowsky2001}. However, results (asymptotic properties) for the $\alpha$--$k$--$NN$ regression is much harder to derive due to the introduction of the power parameter $\alpha$ and are not considered here.

\subsection{Cross-Validation protocol to select the values of $\alpha$ and $k$} \label{sub:sim1}
The 10-fold Cross-Validation (CV) protocol is utilised to tune the pair ($\alpha$, $k$). In the 10-fold CV pipeline, the data are randomly split into 10 folds of nearly equal sizes. One fold is selected to play the role of the test set, while the other folds are considered the training set. The regression models are fitted on the training set and their predictive capabilities are estimated using the test set. This procedure is repeated for the 10 folds so that each fold plays the role of the test set. Ultimately, the predictive performance of each regression model is computed from the aggregation of their predictive performances at each fold. Note that while for Euclidean data the criterion of predictive performance is typically the mean squared error, we instead measure the Kulback-Leibler (KL)  divergence from the observed to the predicted compositional vectors as well as the Jensen-Shannon (JS) divergence which, unlike KL, is a metric, to account for the compositional nature of our response data. The KL and JS measures of divergence are given below:
\begin{subequations}
\begin{eqnarray}
\text{KL}\left({\bf y}, \hat{{\bf y}}\right) &=& \sum_{i=1}^Dy_i\log{\frac{y_i}{\hat y}}. \label{kl} \\
\text{JS}\left({\bf y}, \hat{{\bf y}}\right) &=& \sum_{i=1}^D\left( y_i\log{\frac{2y_i}{y_i+\hat{y}_i}} + \hat{y}_i\log{\frac{2\hat{y}_i}{y_i+\hat{y}_i}} \right). \label{js}
\end{eqnarray}
\end{subequations}

\section{Simulation studies} \label{simulations}
Monte Carlo simulation studies were implemented to assess the predictive performance of the proposed $\alpha$--$k$--$NN$ regression compared to the KLD regression, an alternative semi-parametric approach that also allows for zeros.  

Multiple types of relationships between the response and predictor variables are considered and a 10-fold CV protocol was applied for each regression model to evaluate its predictive performance. A second axis of comparison was an assessment of computational cost of the regressions. Using the same scenario as above, the computational efficiency of the $\alpha$--$k$--$NN$ regression was compared to that of the KLD regression.

All computations were carried out on a laptop with Intel Core i5-5300U CPU at 2.3GHz with 16 GB RAM and SSD installed using the \textit{R} package \textit{Compositional} \citep{compositional2022} for all regression models.

\subsection{Predictive performance}
In our simulation study, the values of the one or two predictor variables (denoted by $\bf x$) were generated from a Gaussian distribution with mean zero and unit variance, and were linked to the compositional responses via two functions: a polynomial as well as a more complex segmented function. For both cases, the outcome was mapped onto $\mathbb{S}^{D-1}$ using Equation (\ref{relation}) 
\begin{eqnarray} \label{relation}
y_i=
\left\lbrace
\begin{array}{cc} 
\frac{1} {1 + \sum_{k=2}^De^{f_k\left({\bf x}\right)} }  & i = 1 \\
\frac{ e^{f_i\left({\bf x}\right)} } {1 + \sum_{k=2}^De^{f_k\left({\bf x}\right)} },  & i = 2,...,D
\end{array}
\right\rbrace.
\end{eqnarray}

More specifically, for the simpler polynomial case, the values of the predictor variables were raised to a power (1, 2 or 3) and then multiplied by a vector of coefficients. White noise (${\bf e}_j$) was added as follows
\begin{eqnarray}  \label{linear}
f_i\left({\bf x}_j\right)  = \left({\bf x}_j^{\nu}\right)^T \pmb{\beta}_i + {\bf e}_j, \ i=2,\ldots,D,\ j=1,\ldots,n
\end{eqnarray}
where $\nu=1,2,3$ indicates the degree of the polynomial. The constant terms in the regression coefficients $\pmb{\beta}_i,\ i=2,\ldots,D$ were randomly generated from $N(-3,1)$ whereas the slope coefficients were generated from  $N(2, 0.5)$.

For the segmented linear model case, one predictor variable $x$ was set to range from $-1$ up to $1$ and the $f$ function was defined as 
\begin{eqnarray}  \label{nonlinear}
f_i\left({\bf x}_j\right) = 
\left \lbrace
\begin{array}{c}
x_j^2 \beta_{1i} + e_j,\ \mathrm{when}\ x_j>0 \\
x_j^3 \beta_{2i} + e_j,\ \mathrm{when}\ x_j<0, 
\end{array}
\right.
\end{eqnarray}
where $i=2,\ldots,D$ and $j=1,\ldots,n$.  The regression coefficients $\beta_{1i}$ were randomly generated from a $N(-1, 0.3)$ while the regression coefficients $\beta_{2i}$, $i=2,\ldots,D$, were randomly generated from $N(1,0.2)$. 

The above two scenarios were repeated with the addition of zero values in 20\% of randomly selected compositional vectors. For each compositional vector that was randomly selected, a third of its component values were set to zero and those vectors were normalised to sum to 1. Finally, for all cases, the sample sizes varied between 100 and 1,000 with an increasing step size equal to 50 while the number of components was set equal to $D=\{3, 5, 7, 10\}$. The estimated predictive performance of the regression models was computed using the KL divergence (\ref{kl}) and the JS divergence (\ref{js}). The results, however, were similar so only the KL divergence results are shown. For all examined case scenarios the results were averaged over 100 repeats. 

\begin{figure}[ht]
\centering
\begin{tabular}{cccc}
\multicolumn{4}{c}{\textbf{$\pmb{\alpha}$--$\bf k$--$\bf NN$ regression}} \\
\includegraphics[scale = 0.35, trim = 30 0 30 0]{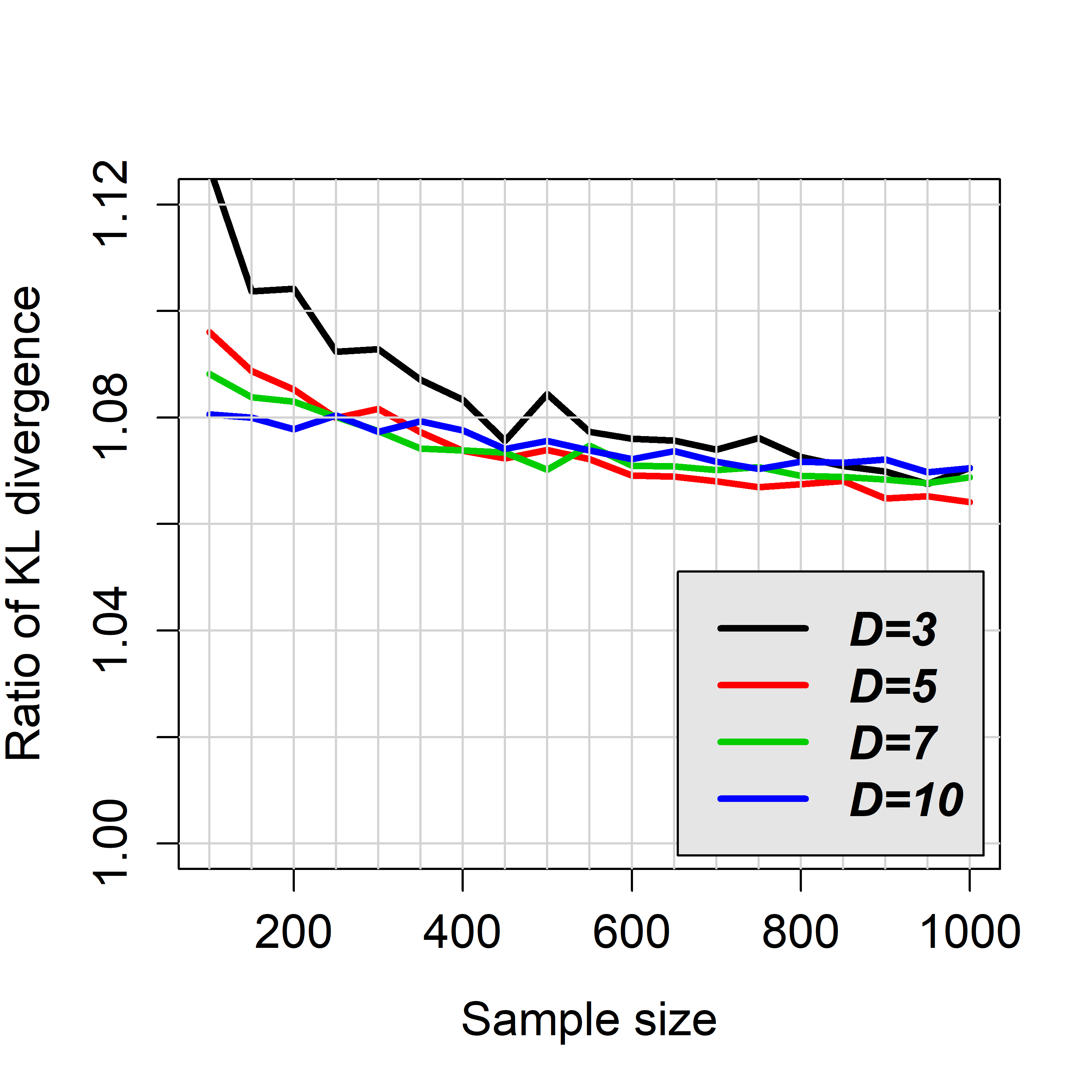}  &
\includegraphics[scale = 0.35, trim = 30 0 30 0]{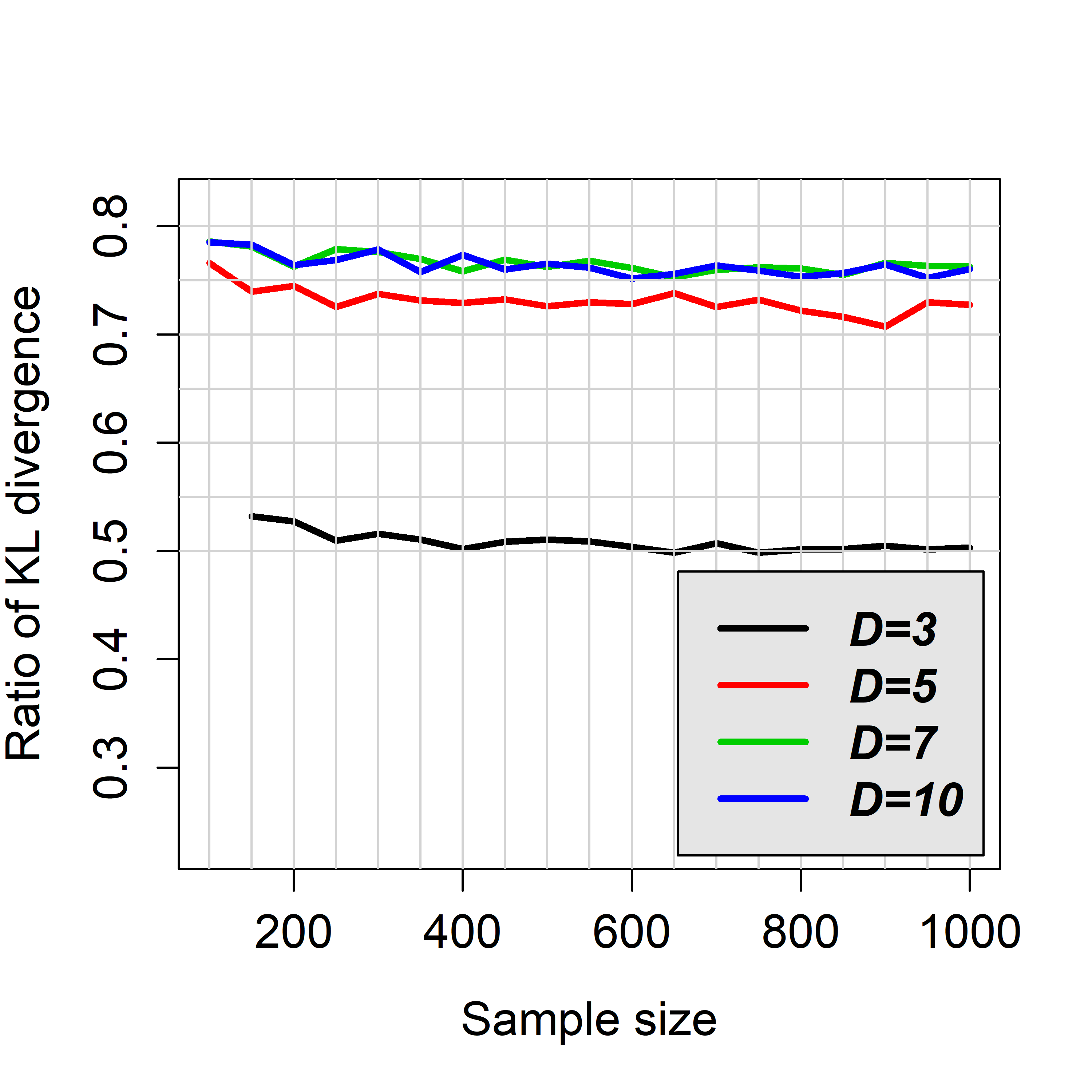}  &
\includegraphics[scale = 0.35, trim = 30 0 30 0]{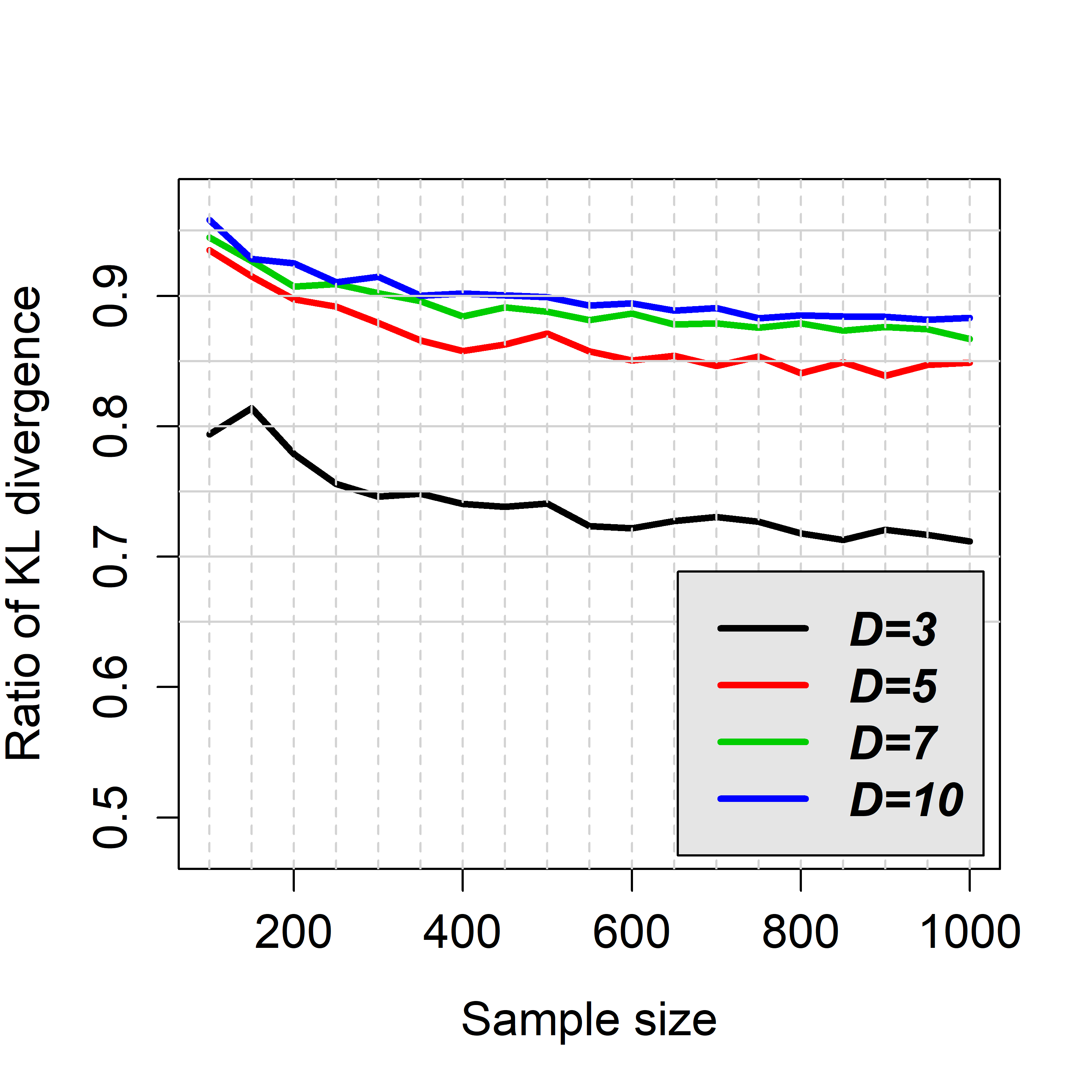}  &
\includegraphics[scale = 0.35, trim = 30 0 30 0]{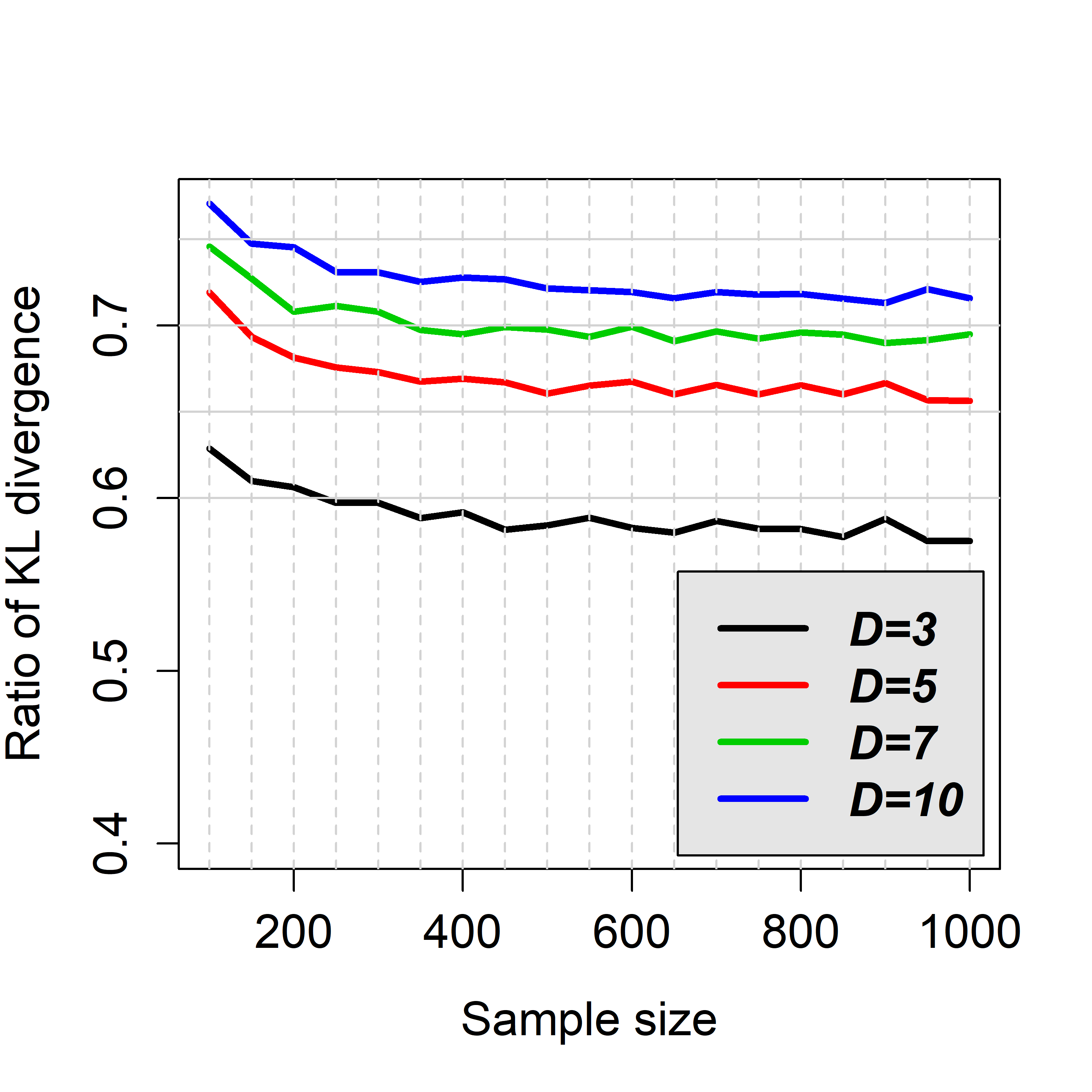}  \\
\textbf{(a)} $\bf \pmb{\nu} = 1$  &  \textbf{(b)} $\bf \pmb{\nu} = 2$  & \textbf{(c)} $\bf \pmb{\nu} = 3$   &  \textbf{(d) Segmented}   
\end{tabular}
\caption{\textbf{No zero values present case scenario}. Ratio of the Kullback-Leibler divergences between the $\alpha$--$k$--$NN$ and the KLD regression. The number of components ($D$) appear with different colors. Values less than $1$ indicate that the $\alpha$--$k$--$NN$ has smaller prediction error than the KLD regression. (a) and (e): The degree of the polynomial in (\ref{linear}) is $\nu=1$, (b) and (f): The degree of the polynomial in (\ref{linear}) is $\nu=2$, (c) and (g): The degree of the polynomial in (\ref{linear}) is $\nu=3$. (d) and (h) refer to the segmented linear relationship case (\ref{nonlinear}).  \label{divergences1} }
\end{figure}

\begin{figure}[ht]
\centering
\begin{tabular}{cccc}
\multicolumn{4}{c}{\textbf{$\pmb{\alpha}$--${\bf k}$--$\bf NN$ regression}} \\
\includegraphics[scale = 0.35, trim = 30 0 30 0]{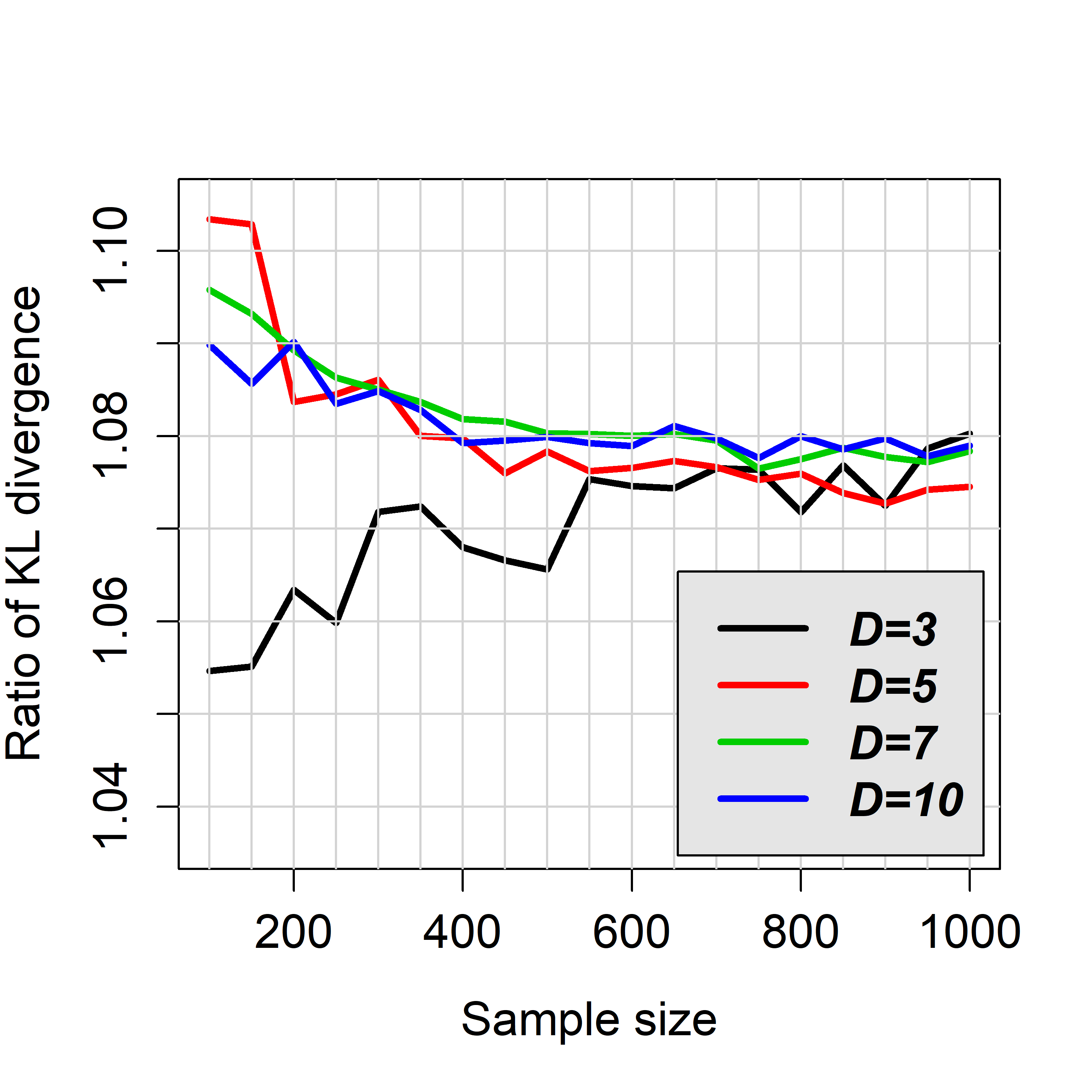}  &
\includegraphics[scale = 0.35, trim = 30 0 30 0]{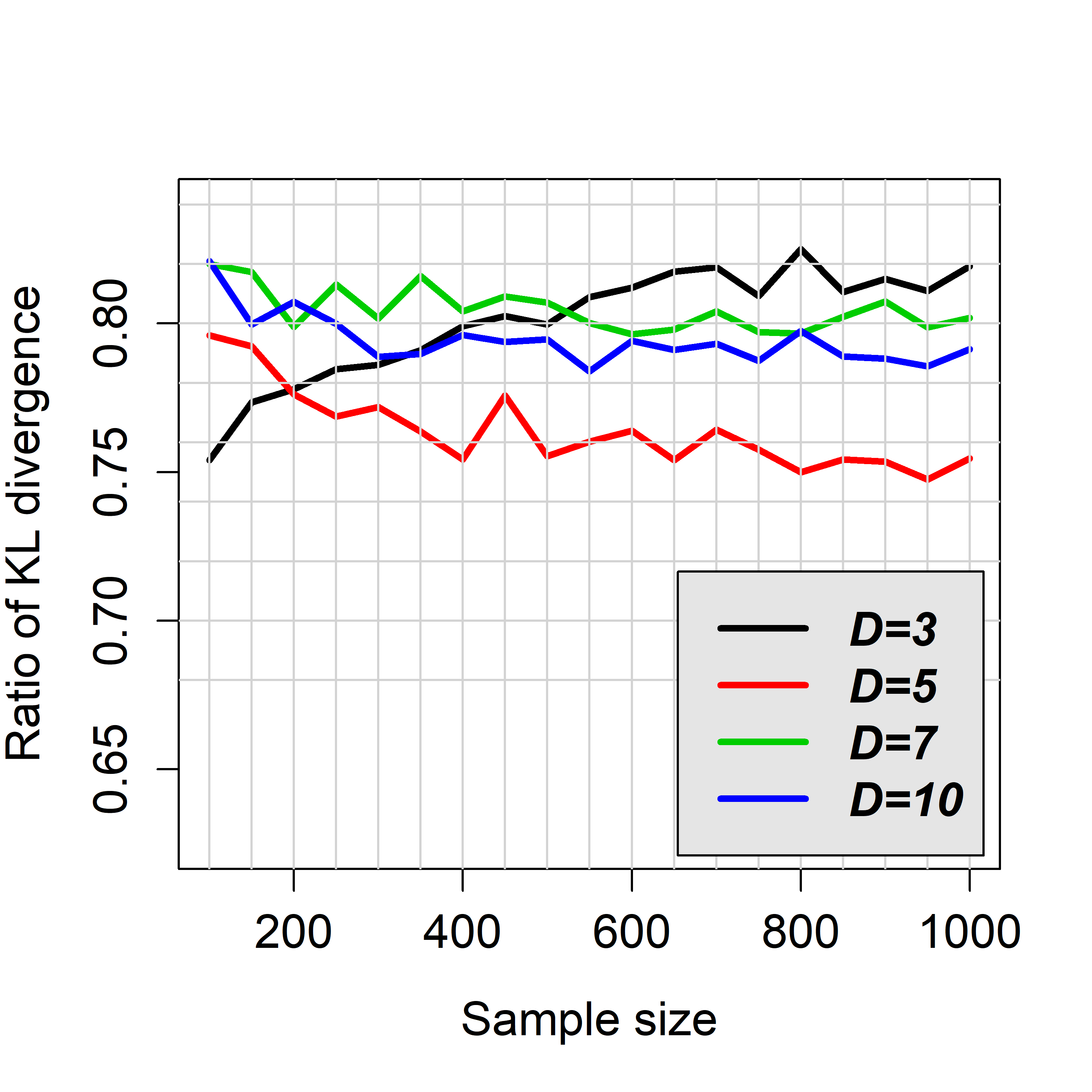}  &
\includegraphics[scale = 0.35, trim = 30 0 30 0]{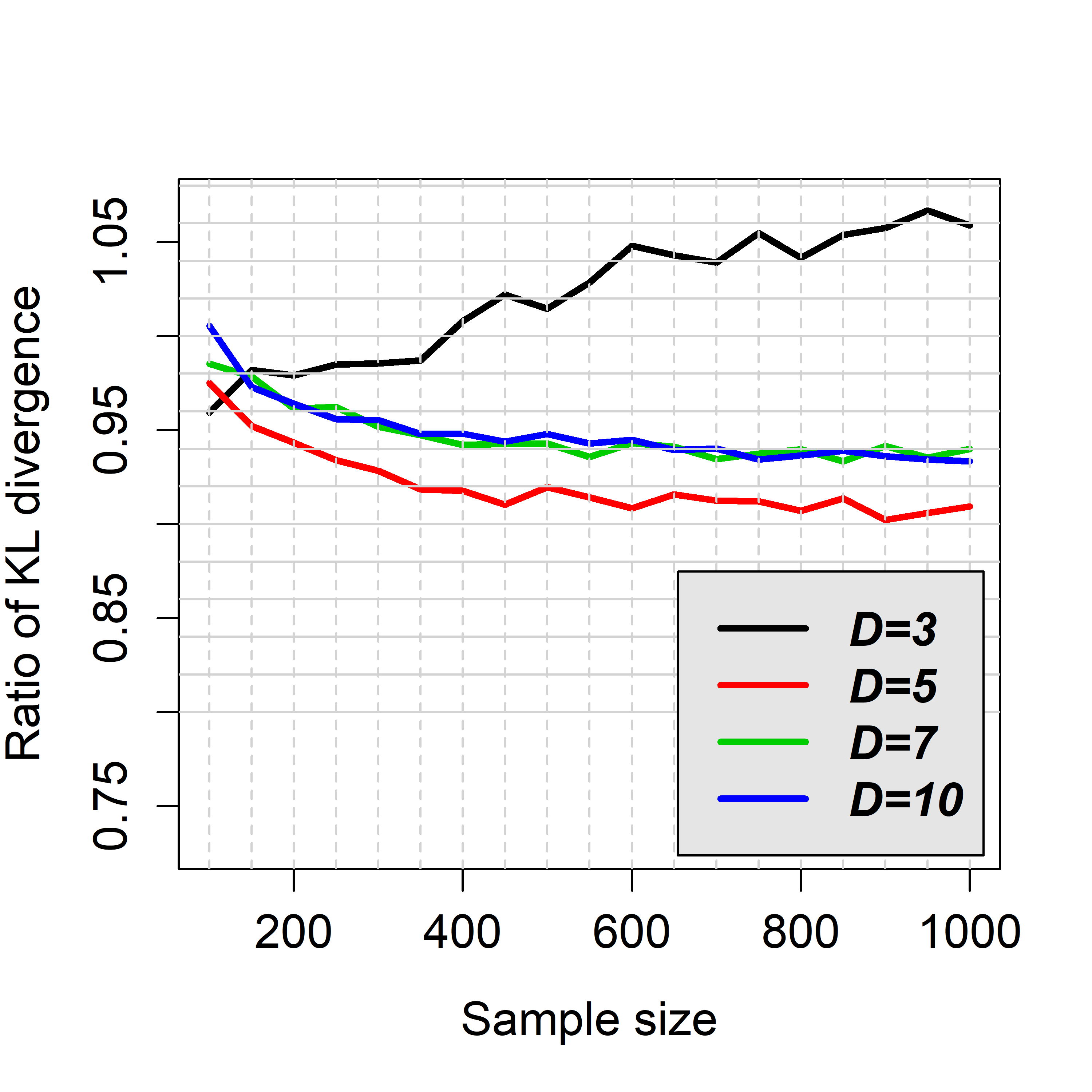}  &
\includegraphics[scale = 0.35, trim = 30 0 30 0]{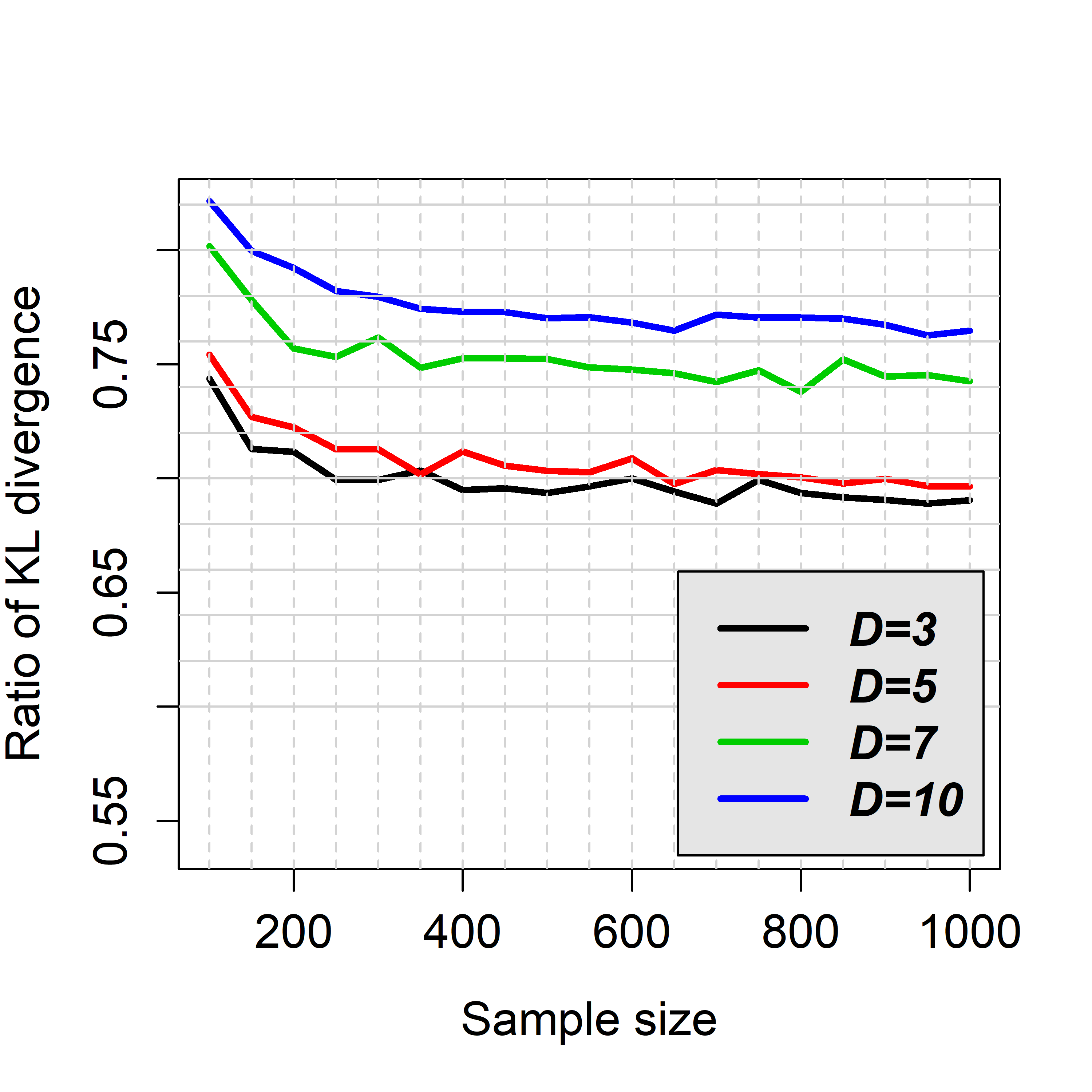}  \\
\textbf{(a)} $\bf \nu = 1$  &  \textbf{(b)} $\bf \nu = 2$  & \textbf{(c)} $\bf \nu = 3$   &  \textbf{(d) Segmented} 
\end{tabular}
\caption{\textbf{Zero values present case scenario}. Ratio of the Kullback-Leibler divergences between the $\alpha$--$k$--$NN$ and the KLD regression. The number of components ($D$) appear with different colors. Values less than $1$ indicate that the $\alpha$--$k$--$NN$ regression has a smaller prediction error than the KLD regression. (a) and (e): The degree of the polynomial in (\ref{linear}) is $\nu=1$, (b) and (f): The degree of the polynomial in (\ref{linear}) is $\nu=2$, (c) and (g): The degree of the polynomial in (\ref{linear}) is $\nu=3$. (d) and (h) refer to the segmented linear relationship case (\ref{nonlinear}).  \label{divergences2} }
\end{figure}

Figures \ref{divergences1} and \ref{divergences2} show graphically the results of the comparison of $\alpha$--$k$--$NN$ regression with KLD regression with no zeros and zero values present, respectively. Note that values below $1$ indicate that the proposed regression models have smaller prediction errors than the KLD regression. For the first case of no zero values present (Figure \ref{divergences1}), when the relationship between the predictor variable(s) and the compositional responses is linear ($\nu=1$), the error is slightly less for KLD regression compared to $\alpha$--$k$--$NN$ regression. In all other cases (quadratic, cubic and segmented relationships), the $\alpha$--$k$--$NN$ regression consistently produces more accurate predictions. Another characteristic observable in all plots of Figure \ref{divergences1} is that the relative predictive performance of both non-parametric regressions compared to KLD regression reduces as the number of components in the compositional data increases. 

The results in the zero values present case in Figure \ref{divergences2} are, in essence, the same compared to the previous case. When the relationship between the predictor variables is linear, KLD regression again exhibits slightly more accurate predictions compared to the $\alpha$--$k$--$NN$ regression, while the opposite is true for most other cases. Furthermore, the impact of the number of components of the compositional responses on the relative predictive performance of both non-parametric regressions compared to the KLD regression varies according to the relationship between the response and covariates, the number of predictor variables and the sample size.

\subsection{Computational efficiency of the $\alpha$--$k$--$NN$ regression}
The linear relationship scenario (that is, when the degree of the polynomial in Equation (\ref{linear}) is equal to $\nu=1$), but without zero values was used to illustrate the computational efficiency of only the $\alpha$--$k$--$NN$ regression. 

With massive data (tens or hundreds of thousands of observations) the search for the $k$ nearest neighbours in the $\alpha$--$k$--$NN$ regression takes place using kd-trees implemented in the \textit{R} package \textit{RANN} \citep{rann2019}. An important advantage of a kd-tree is that it runs in $O(n\log{n})$ time, where $n$ is the sample size of the training data set. The \textit{RANN} package utilizes the Approximate Near Neighbor (ANN) C++ library, which can give the exact near neighbours or (as the name suggests) approximate near neighbours to within a specified error bound, but when the error bound is $0$ (as in this case) an exact nearest neighbour search is performed. When the sample sizes are at the order of a few tens of thousands or less, kd-trees are slower to use and, in this case, the $\alpha$--$k$--$NN$ regression algorithm employs a C++ implemented function to search for the $k$ nearest neighbours from the \textit{R} package \textit{Rfast} \citep{rfast2022}. 

In this study, the number of components of the compositional data and the number of predictor variables were kept the same as before but the sample sizes, the values of $\alpha$ and the number of neighbours were modified. Specifically, large sample sizes ranging from $500,000$ up to $10,000,000$ with an increasing step equal to $500,000$ were generated. Eleven positive values of $\alpha$ ($0, 0.1,\ldots, 1$) were used and a large sequence of neighbours (from $k=2$ up to $k=100$ neighbours) were considered. The computational efficiency of each regression was measured as the time required to predict the compositional responses of $1,000$ new values. 

The average time (in seconds) based on 10 repetitions versus the sample size of each regression is presented in Tables \ref{tab_time1} and \ref{tab_time2}. The scalability of $\alpha$--$k$--$NN$ regression is better than that of KLD regression and as the sample size explodes the difference in the computational cost increases. Furthermore, the ratio of the computational cost of $\alpha$--$k$--$NN$ regression to the cost of OLS decays with the sample size and with the number of components. The converse is true for the ratio of the computational cost of KLD to the cost of OLS with respect to the number of components which appears to increase with the number of components. To appreciate the level of computational difficulty it should be highlighted that the $\alpha$--$k$--$NN$ regression produced a collection of $11 \times 99 = 1089$ predicted compositional data sets (for each combination of $\alpha$ and $k$). KLD, in contrast, produced a single set of predicted compositional data. The same is true for OLS regression whose time required for the same task is also presented.

\begin{table}[!ht]
\caption{Computational cost of the regression with no zero values present. Time in seconds, of the each regression method for various sample size required by each regression for various sample sizes and $D=3$ and $D=5$ components. Inside the parentheses appear the ratios of the computational cost of each method compared to the computational cost of OLS. \label{tab_time1}}
\begin{center}
\begin{tabular}{l|ccc|ccc}  \hline \hline
              & \multicolumn{3}{c}{$D=3$}  &  \multicolumn{3}{c}{$D=5$}  \\   \hline \hline
Sample size   & OLS  & $\alpha$--$k$--$NN$  &  KLD  & OLS  & $\alpha$--$k$--$NN$  &  KLD   \\ \hline
$n = 1\times10^6$    &  0.18  &  3.13(17.10)   &  3.17(17.34)   &  0.32  &  3.99(12.40)  &  7.77(24.14)   \\
$n = 2\times10^6$    &  0.40  &  4.86(12.15)   &  7.34(18.36)   &  0.78  &  4.89(6.29)   &  14.57(18.75)  \\
$n = 3\times10^6$    &  0.64  &  7.91(12.40)   &  12.72(19.94)  &  1.14  &  6.46(5.65)   &  21.70(18.99)  \\
$n = 4\times10^6$    &  0.80  &  10.93(13.57)  &  16.84(20.92)  &  1.29  &  8.60(6.65)   &  31.20(24.17)  \\
$n = 5\times10^6$    &  1.35  &  11.65(8.62)   &  17.49(12.94)  &  1.58  &  13.13(8.30)  &  43.71(27.63)  \\
$n = 6\times10^6$    &  1.68  &  15.87(9.46)   &  22.44(13.38)  &  2.25  &  12.34(5.48)  &  43.25(19.21)  \\
$n = 7\times10^6$    &  2.14  &  17.91(8.38)   &  27.23(12.75)  &  2.19  &  15.90(7.24)  &  47.83(21.79)  \\
$n = 8\times10^6$    &  2.10  &  21.92(10.45)  &  28.76(13.71)  &  2.54  &  17.71(6.98)  &  60.49(23.83)  \\
$n = 9\times10^6$    &  2.34  &  19.18(8.18)   &  29.34(12.52)  &  2.94  &  20.77(7.06)  &  70.96(24.13)  \\
$n = 10\times10^6$   &  2.45  &  19.79(8.07)   &  31.92(13.01)  &  3.28  &  21.47(6.55)  &  80.02(24.41)  \\  \hline \hline
\end{tabular}
\end{center}
\end{table}

\begin{table}[!ht]
\caption{Computational cost of the regression with no zero values present. Time in seconds, of the each regression method versus the sample size required by each regression for various sample sizes and $D=7$ and $D=10$ components. Inside the parentheses appear the ratios of the computational cost of each method compared to the computational cost of OLS. \label{tab_time2}}
\begin{center}
\begin{tabular}{l|ccc|ccc}  \hline \hline
              & \multicolumn{3}{c}{$D=7$}  &  \multicolumn{3}{c}{$D=10$}  \\   \hline \hline
Sample size   & OLS  & $\alpha$--$k$--$NN$  &  KLD  & OLS  & $\alpha$--$k$--$NN$  &  KLD   \\ \hline
$n = 1\times10^6$   &  0.50  &  4.11(8.19)   &  12.06(24.02)   &  0.57  &  4.78(8.38)    &  20.98(36.81)    \\
$n = 2\times10^6$   &  0.91  &  5.51(6.03)   &  24.35(26.64)   &  1.28  &  6.74(5.28)    &  48.88(38.30)    \\
$n = 3\times10^6$   &  1.47  &  7.57(5.15)   &  36.65(24.95)   &  2.06  &  9.58(4.65)    &  76.62(37.19)    \\
$n = 4\times10^6$   &  1.72  &  8.11(4.71)   &  44.51(25.83)   &  2.19  &  8.23(3.75)    &  78.63(35.86)    \\
$n = 5\times10^6$   &  2.11  &  8.94(4.24)   &  53.77(25.50)   &  3.02  &  10.13(3.35)   &  102.63(33.93)   \\
$n = 6\times10^6$   &  3.13  &  11.17(3.57)  &  65.74(21.00)   &  3.54  &  12.91(3.64)   &  133.86(37.78)   \\
$n = 7\times10^6$   &  3.44  &  14.46(4.20)  &  82.31(23.90)   &  4.40  &  15.20(3.46)   &  171.33(38.95)   \\
$n = 8\times10^6$   &  3.53  &  18.03(5.11)  &  108.84(30.87)  &  5.15  &  17.31(3.36)   &  199.97(38.84)   \\
$n = 9\times10^6$   &  4.13  &  21.37(5.18)  &  117.37(28.43)  &  7.77  &  23.61(3.04)   &  263.47(33.92)   \\
$n = 10\times10^6$  &  4.90  &  23.56(4.80)  &  139.84(28.51)  &  8.00  &  24.34(3.04)   &  312.00(38.98)   \\  \hline \hline
\end{tabular}
\end{center}
\end{table}

\section{Examples with real data} \label{data}

\subsection{Small sample sized data sets}
To assess the predictive performance of $\alpha$--$k$--$NN$ regression in practice, 7 publicly available small sample sized data sets, with compositional responses, were utilised as examples. The same 10-fold CV protocol as before (see Subsection \ref{sub:sim1}) was repeated using the 7 real data sets which are described briefly below. Note that the names of the data sets are consistent with the names previously used in the literature. A summary of the characteristics of the data sets, including the dimension of the response matrix, the number of compositional response vectors containing at least one zero value as well as the number of predictor variables, is provided in Table \ref{tab_real_1}.

\begin{itemize}
\item \textbf{Lake}: Measurements in silt, sand and clay were taken at 39 different water depths in an Arctic lake. The question of interest was to predict the composition of these three elements for a given water depth. The data set is available in the \textit{R} package \textit{compositions} \citep{compositions2018} and contains no zero values.
\item \textbf{Glacial}: In a pebble analysis of glacial tills, the percentages by weight in 92 observations of pebbles of glacial tills sorted into 4 categories were recorded. The glaciologist was interested in predicting the compositions based on the total pebbles counts. The data set is available in the \textit{R} package \textit{compositions} \citep{compositions2018} and almost half of the observations (42 out of 92) contain at least one zero value.
\item \textbf{GDP}: The 2009 GDP per capita of the 27 member states of the European Union as well as the mean household consumption expenditures (in Euros) in 12 categories are provided by  \href{https://ec.europa.eu/eurostat/home?}{eurostat}. The data are available in \cite{egozcue2012} and they contain no zero values.
\item \textbf{Gemas}: This data set contains 2083 compositional vectors containing the concentration in 22 chemical elements (in mg/kg). The data set is available in the \textit{R} package \textit{robCompositions} \citep{robcompositions2011} with 2108 vectors, but 25 vectors had missing values and thus were excluded from the current analysis. There was only one vector with one zero value. The predictor variables are the annual mean temperature and annual mean precipitation.
\item \textbf{Fish}: This data set provides information on the mass (the only predictor variable) and 26 morphometric measurements (the compositional response data) for 75 Salvelinus alpinus (a type of fish). The data set is available in the \textit{R} package \textit{easyCODA} \citep{greeacre2018} and contains no zero values.
\item \textbf{Data}: In this data set, the compositional response is a matrix of 9 party vote-shares across 89 different democracies (countries) and the (only) predictor variable is the average number of electoral districts in each country. The data set is available in the \textit{R} package \textit{compositions} \citep{ocomposition2015} and 80 out of the 89 vectors contain at least one zero value.
\item \textbf{Elections}: The Elections data set contains information on the 2000 U.S. presidential election in the 67 counties of Florida. The number of votes each of the 10 candidates received was transformed into proportions. For each county, information on 8 predictor variables was available. The data set is available in \cite{smith2002} and 23 out of the 67 vectors contained at least one zero value. 
\end{itemize}

\begin{figure}[ht]
\centering
\begin{tabular}{cc}
\includegraphics[scale = 0.4]{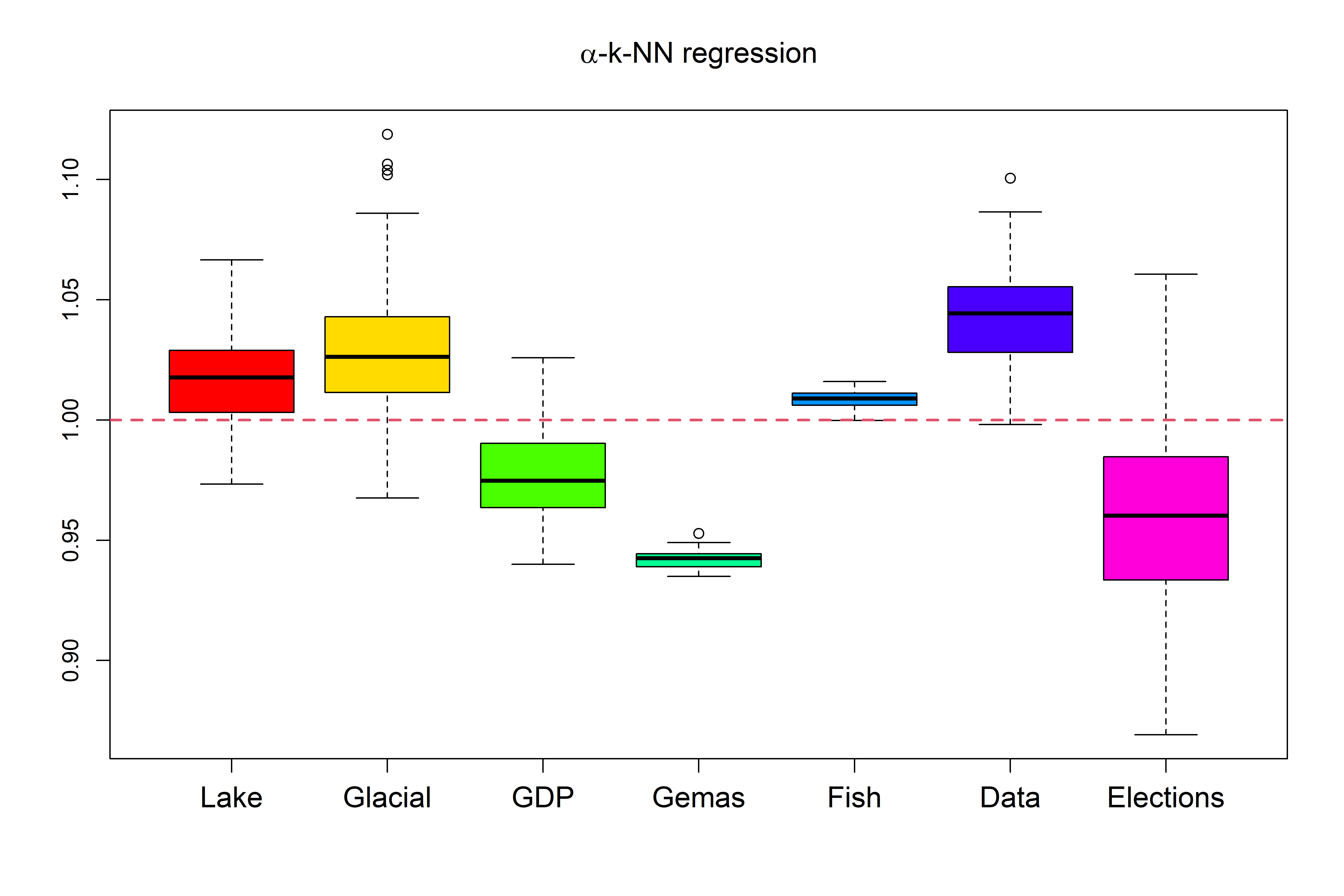}  \\
\includegraphics[scale = 0.4]{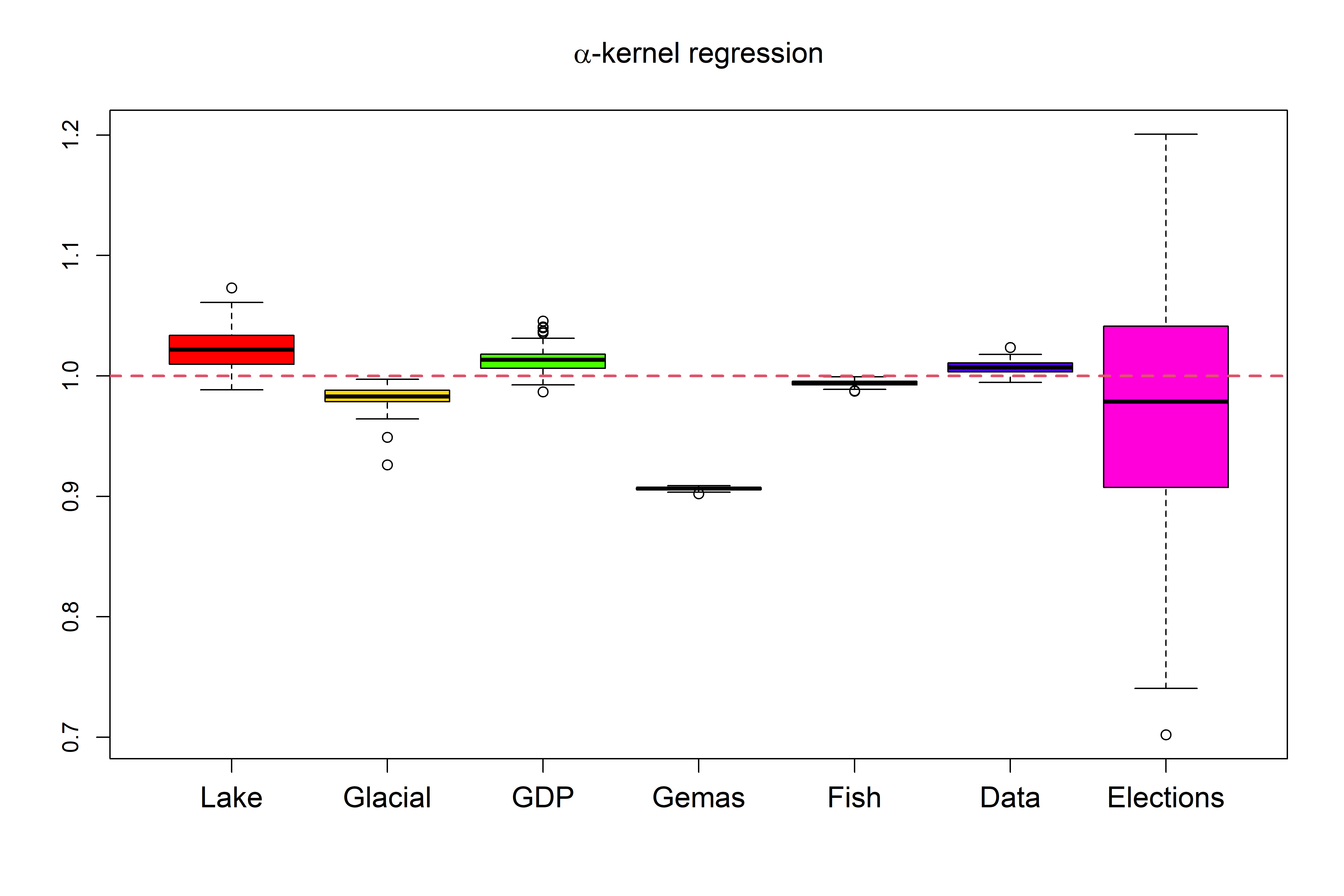}  
\end{tabular}
\caption{Ratio of the Kullback-Leibler divergence between the $\alpha$--$k$--$NN$ regression to the KLD regression. Values lower than $1$ indicate a smaller prediction error compared to the KLD regression. \label{divergences} }
\end{figure}

Figure \ref{divergences} presents the boxplots of the relative performance (computed via the KL divergence) of $\alpha$--$k$--$NN$ regression compared to KLD regression for each data set. As before, values lower than $1$ indicate that the proposed regression algorithm has smaller prediction error than KLD regression. On average, the $\alpha$--$k$--$NN$ outperformed the KLD regression for the data sets \textbf{GDP}, \textbf{Gemas} and \textbf{Elections}, whereas the opposite was true for the data set \textbf{Lake}, \textbf{Glacial}, \textbf{Fish} and \textbf{Data}.

Table \ref{tab_real_1} presents the most frequently selected values of $\alpha$ and $k$. It is perhaps worth mentioning that the value of $\alpha=0$ was never selected for any data set, indicating that the ilr transformation in Equation (\ref{ilr}) was never considered the optimal transformation. When the percentage of times the value of $\alpha$ is selected is large, this implies a small variance in the chosen value of the parameter. From Table \ref{tab_real_1}, it appears that the larger the value of the optimal $\alpha$ (roughly), the smaller the variance. For example, for the data set \textbf{Lake}, the optimal value $\alpha$ was chosen to be $1$, 93\% of the time for the $\alpha$--$k$--$NN$ regression. For the data set \textbf{Fish}, however, the optimal value of $\alpha$, as well as the percentage of time it was chosen, were smaller for both regressions. There does not appear to be an association between the variability in the chosen value of $k$ and the variability in the optimal $\alpha$ values for the $\alpha$--$k$--$NN$ regression. For example, for both data sets \textbf{Lake} and \textbf{Fish}, 10 nearest neighbours were chosen only 66\% of the time. For the data set \textbf{Gemas}, the choice of $\alpha$ was highly variable, whereas the choice of $k$ was always the same. The opposite was true for the data set \textbf{Data}, for which the optimal value of $\alpha$ was always the same but the value of $k$ was highly variable.

\begin{table}[!ht]
\caption{Information about the response compositional variables, the predictor variables and the pairs ($\alpha$, $k$) chosen for every data set, including the most frequently selected value and the percentage of times it is selected. \label{tab_real_1}}
\begin{center}
\begin{tabular}{l|l|c|c|r|r}  \hline \hline
     &      &           &   &  \multicolumn{2}{c}{$\alpha$--$k$--$NN$}  \\
Data Set &  Dimensions  & No of vectors  & No of & \% of times $\alpha$ & \% of times $k$   \\ 
&  ($n \times D$)  &  with zero values  & predictors  &  was selected  &  was selected   \\  \hline \hline
\textbf{Lake}      &  $39 \times 3$   &  0   & 1  &  1 (93\%)    & 10 (66\%)     \\        
\textbf{Glacial}   & $92 \times 4$    &  42  & 1  &  1 (93\%)    & 10 (75\%)     \\
\textbf{GDP}       &  $27 \times 12$  &  0   & 1  &  0.8 (33\%)  & 5 (89\%)     \\
\textbf{Gemas}     & $2083 \times 30$ &  1   & 2  &  0.5 (47\%)  & 10 (100\%)    \\
\textbf{Fish}      & $75 \times 26$   &  0   & 1  &  -1 (29\%)   & 20 (66\%)     \\
\textbf{Data}      & $89 \times 9$    & 80   & 1  &  1 (100\%)   & 10 (28\%)     \\
\textbf{Elections} & $67 \times 10$   & 23   & 8  &  0.5 (30\%)  & 7 (32\%)     \\  \hline \hline
\end{tabular}
\end{center}
\end{table}

\begin{figure}[ht]
\centering
\begin{tabular}{ccc}
\includegraphics[scale = 0.33, trim = 40 0 0 0]{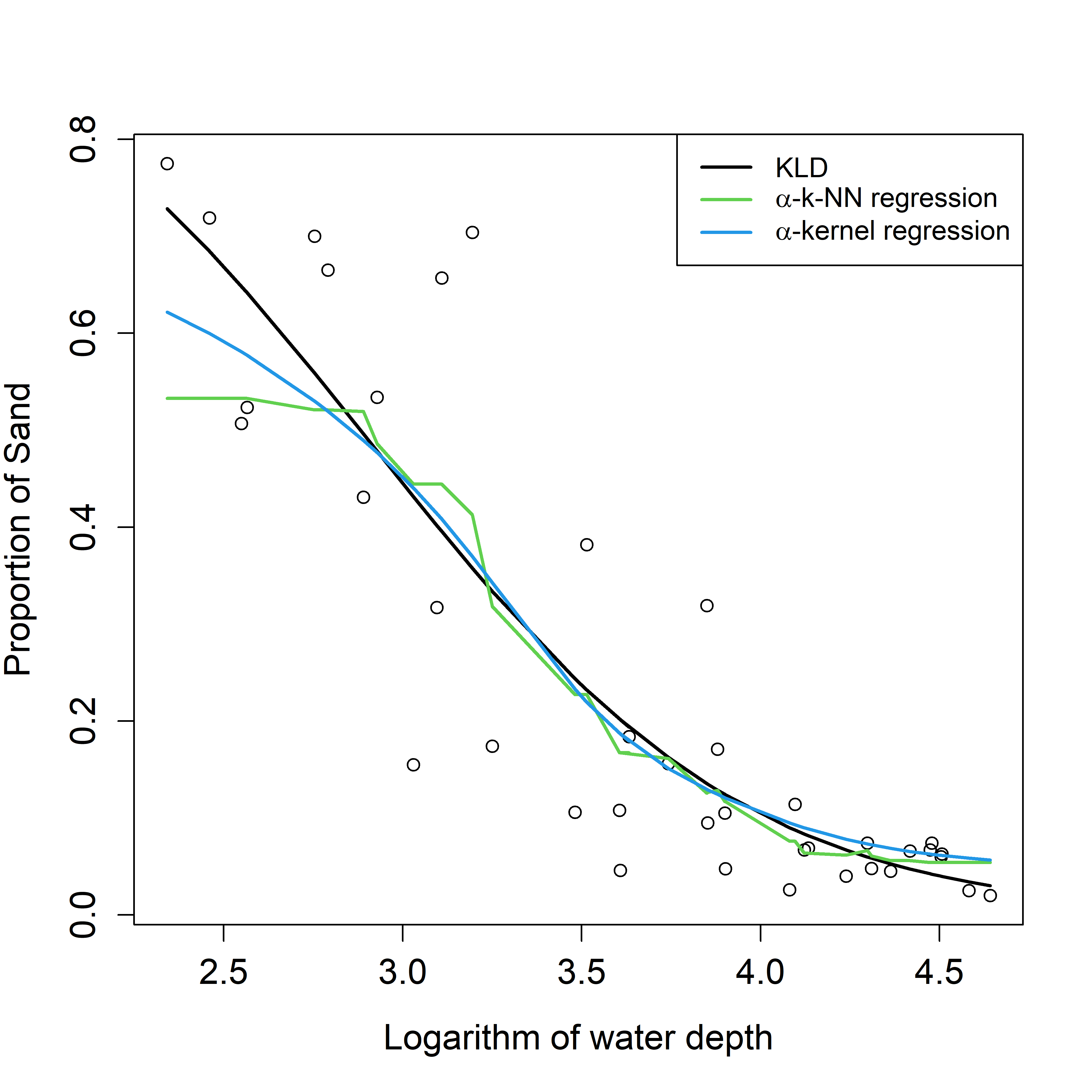}  &
\includegraphics[scale = 0.33, trim = 50 0 0 0]{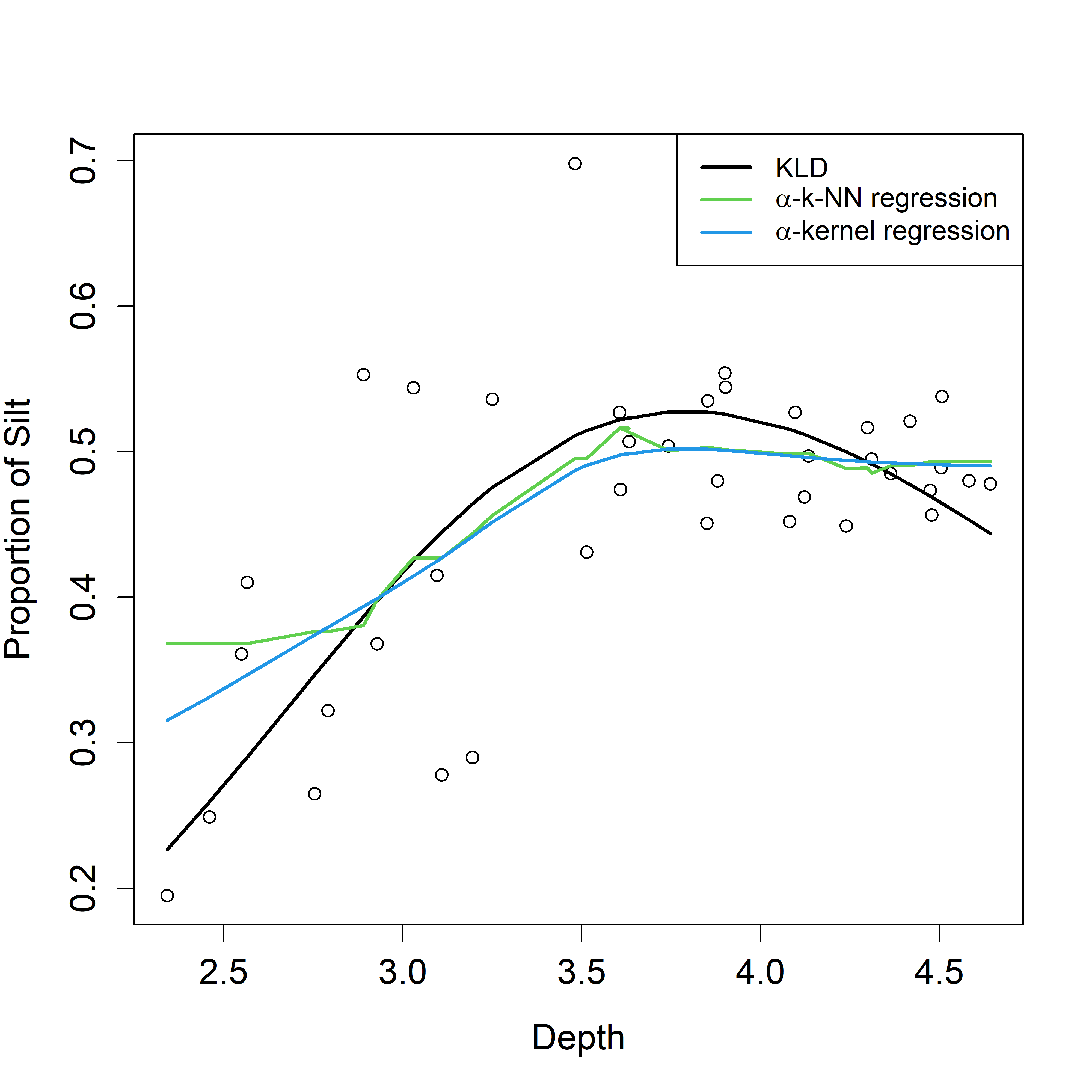}  & 
\includegraphics[scale = 0.33, trim = 50 0 0 0]{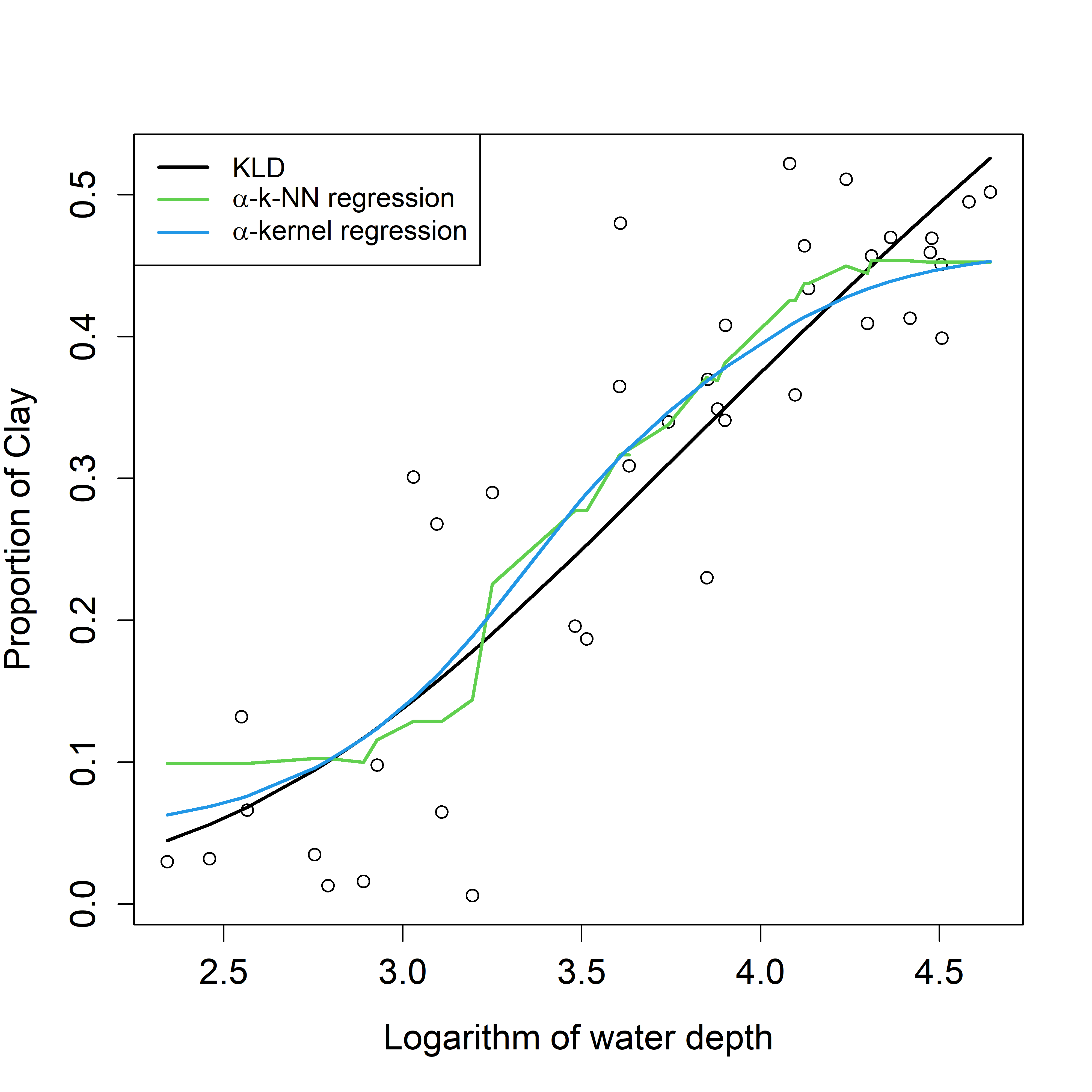}  \\
(a)  &  (b) & (c)
\end{tabular}
\caption{Arctic lake data: the logarithm of the water depth against the observed fitted values based on the three regression models. \label{arctic} }
\end{figure}

\begin{figure}[ht]
\centering
\includegraphics[scale = 0.33]{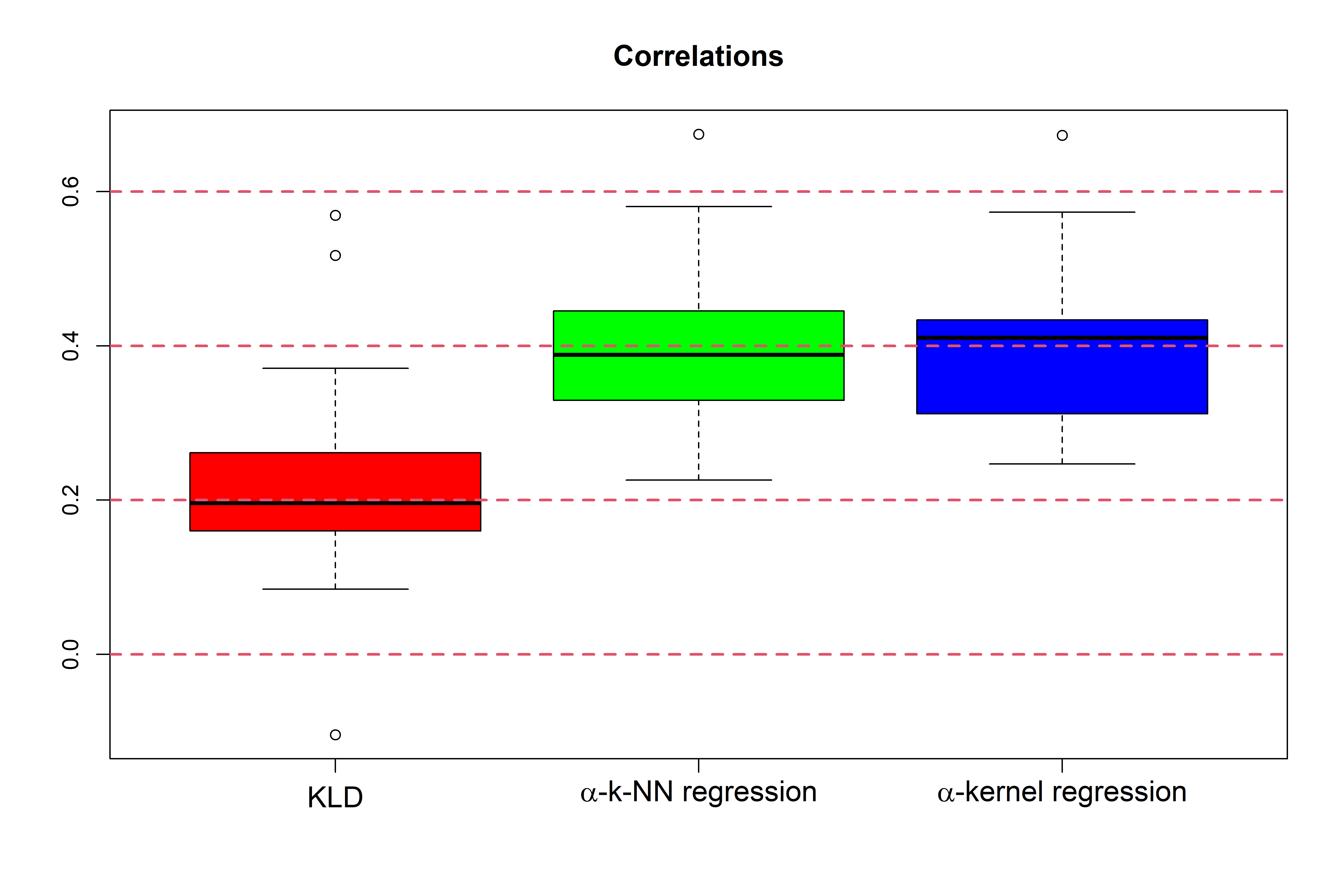} 
\caption{Arctic lake data: boxplot of the correlations between the observed and fitted values of each component for the three regression models. \label{gemas} }
\end{figure}

Evidently, as the dimensions increase visual inspection of the fit of the models becomes harder. To bypass this problem the correlations between each component observed and fitted values may be computed. Figure \ref{gemas} presents the boxplots of the correlations for each model under examination, where it is evident that the non-parametric models have superseded the KLD regression model.

The results from the real data analysis do not come by surprise and in fact corroborate the findings of the simulation studies. The simulation studies showed that when the relationship between the compositional responses and the independent variable(s) is linear, the $\alpha$--$k$--$NN$ regression model does not offer an improved fit over their competitor, the KLD regression. But, when the relationship is not linear the non-parametric models are to be preferred. For the Arctic lake data for example, the correlations between the alr transformed compositional data and the logarithm of the water depth are high and a scatter plot clearly shows a highly linear relationship. For the Gemma data set though, the correlations between the alr transformed compositional data and the two independent variables are rather low. Hence, this could be used as a rule of thumb for initial inspection as to whether the KLD regression should be selected against a non-parametric regression model.

\subsection{Large scale sample sized data sets}
A benefit of $\alpha$--$k$--$NN$ regression is its high computational efficiency and hence we also illustrate its performance on two real large scale data sets. 

\begin{itemize}
\item \textbf{Seoul pollution}: Air pollution measurement information in Seoul, South Korea, is provided by the Seoul Metropolitan Government `Open Data Plaza'. This particular data set was downloaded from \href{https://www.kaggle.com/bappekim/air-pollution-in-seoul}{kaggle}. The average values for 4 pollutants are available along with the coordinates (latitude and longitude) of each site. The data were normalised to sum to 1, so as to obtain the composition of each pollutant. In total, there are 639,073 of observations. Since the predictors in the \textbf{Seoul pollution} data set, the longitude and latitude, are expressed in polar coordinates they were first transformed to their Euclidean coordinates, using the \textit{R} package \textit{Directional} \citep{directional2022}, in order to validly compute the Euclidean distances.      
\item \textbf{Electric power consumption}: This data set contains 1,454,154 measurements of electric power consumption in one household with a one-minute sampling rate over a period of almost 4 years. The measurements were gathered in a house located in Sceaux between December 2006 and November 2010 (47 months). Different electrical quantities and some sub-metering values are available. The data set is available to download from the \href{https://archive.ics.uci.edu/ml/datasets/Individual+household+electric+power+consumption}{UCI Machine Learning Repository}. The response data comprised of 3 energy sub meterings measured in watt-hour of active energy. The data were again transformed to compositional data. There are 4 predictor variables.
\end{itemize}

The same 10-fold CV protocol was employed again. Since both compositional data sets contained zero values,the KLD and $\alpha$--$k$--$NN$ regression methods were suitable. Only strictly positive values of $\alpha$ were therefore utilised, and for the nearest neighbours, 99 values were tested ($k=2,\ldots,100$). The results of the $\alpha$--$k$--$NN$ and the KLD regression are summarised in Table \ref{tab_real_2}. 

\begin{table}[!ht]
\caption{The minimum Kullback-Leibler and Jensen-Shannon divergence obtained by the $\alpha$--$k$--$NN$ and the KLD regression for each large scale data. The optimal combination of $\alpha$ and $k$ appears inside the parentheses. \label{tab_real_2}}
\begin{center}
\begin{tabular}{l|cc|cc}  \hline \hline
Data Set  &  \multicolumn{2}{c}{Kullbak-Leibler divergence}  &  \multicolumn{2}{c}{Jensen-Shannon divergence}   \\  \hline \hline
          &  $\alpha$--$k$--$NN$  & KLD    & $\alpha$--$k$--$NN$  & KLD  \\  \hline      
\textbf{Seoul Polution}   & 0.068 ($\alpha$=1, $k$=67) &  0.047  & 0.005 ($\alpha$=1, $k$=67)  &  0.004   \\        
\textbf{Electric power consumption}   & 0.541 ($\alpha$=1, $k$=2) &  0.825  & 0.046 ($\alpha$=1, $k$=2)  &  0.063  \\ \hline \hline
\end{tabular}
\end{center}
\end{table}

These two large scale data sets suggest that, similarly to the small scale data sets, $\alpha$--$k$--$NN$ is a viable alternative regression model option for compositional data, with the advantage that it is as computationally efficient, or more than, other regression models, at the cost of interpretability of the effects of the predictor variables. 

\section{Conclusions} \label{conclusion}
Two generic regressions able to capture complex relationships involving compositional data, termed $\alpha$--$k$--$NN$ regression was proposed that take into account the constraints on such data. Through simulation studies and the analysis of several real-life data sets, $\alpha$--$k$--$NN$ regression were evaluated alongside a comparable, but semi-parametric, regression model available for this setting. The classical $k$--$NN$ regression provided the foundation for $\alpha$--$k$--$NN$ regression, while the $\alpha$--transformation was used to transform the compositional data. Using this transformation added to the flexibility of the model and meant that commonly occurring zero values in the compositional data were allowed, unlike with many other regression approaches for compositional data. The Fr{\'e}chet mean (defined for compositional data by \cite{tsagris2011}) was used in order to prevent fitted values from being outside the simplex.

Using a CV procedure and pertinent measures of predictive performance, we found that in simulation study cases where the relationship was non-linear (including when the data contained zeros), the $\alpha$--$k$--$NN$ regression outperformed (sometimes substantially) their competing counterpart, KLD regression. For the real-life data sets, similar conclusions were made and our two non-parametric regressions tended to outperform KLD regression in data sets where it is surmised that non-linear relationships exist. We note that our conclusions were the same regardless of which type of divergence was used (either KL or JS). A second advantage of the $\alpha$--$k$--$NN$ regression solely, is its high computational efficiency as it can treat millions of observations in just a few seconds. We further highlight that the new regression techniques are publicly available in the \textit{R} package \textit{Compositional} \citep{compositional2022}.

A disadvantage of the $\alpha$--$k$--$NN$ regression, and of $k$--$NN$ regression in general, is that it lacks the framework for classical statistical inference (such as hypothesis testing). This is counterbalanced by a) its higher predictive performance compared to parametric models and b) its high computational efficiency that make it applicable even with millions of observations. However, the use of ICE plots \citep{goldstein2015} that offer a visual inspection of the effect of each independent variable can overcome this issue. Note that while not considered in detail here, $\alpha$-regression \citep{tsagris2015b}  can also handle zeros but is computationally expensive.

\end{document}